\documentclass{ametsocV6.1}
\usepackage{makecell}
\usepackage[table]{xcolor}
\usepackage{mathtools}
\usepackage{placeins}
\usepackage{natbib}

\title{Idealized Cumulus Cloud-Scale Motions and the Dynamics of Isolated and Coupled Flows
}

\authors{Dario Falcone,\aff{a}\correspondingauthor{Dario Falcone, dfalcone@ucdavis.edu} 
Matthew R. Igel,\aff{b} 
Joseph A. Biello,\aff{c}}

\affiliation{\aff{a}{Atmospheric Science Graduate Group, University of California Davis}\\
\aff{b}{Department of Land, Air and Water Resources, University of California Davis}\\
\aff{c}{Department of Applied Mathematics, University of California Davis}}

\abstract{Developing an understandable theory for the dynamic evolution of the morphology of clouds remains intractable.  To break this deadlock, we introduce a new conceptual model for cloud-scale motions named the Kinematics Representation of Non-rotating Updraft Tori (KRoNUT) model, where non-rotating reflects the absence of motion in the azimuthal direction.  Using this model, we conduct a series of relaxation experiments whereby we ``turn off'' the baroclinic term associated with a pre-existing cloud-scale circulation. We then implement a moment reduction technique to generate a system of differential equations named the Dynamics of Non-rotating Updraft Tori (DoNUT) equations, which describe the temporal evolution of a cloudy circulation under various combinations of forcings,  namely turbulent diffusion, self-advection, and cross-advection from a neighboring cloud-scale flow.  The solutions of the DoNUT equations show that all single KRoNUT configurations either start at or evolve toward a specific steady state circulation.  The cloud-scale motions represented by the current KRoNUT model always grow vertically but may narrow, due to advection, or widen, due to diffusion. Meanwhile, invigoration or enervation of the vertical velocity may result from advection or diffusion processes, with short, wide KRoNUTs more likely to invigorate and tall, narrow KRoNUTs likely to enervate.  Our study of the coupled KRoNUTs provides insight into clouds' tendencies to attract or repel one another. Important results of the coupled KRoNUT analysis include a scaled metric for interaction, ranges of specific height ratios that induce the most meaningful interaction, and circulation parameters that alter the location and stability of a steady KRoNUT.}

\begin{document}

\maketitle

%
%
%
%
%
%
\statement Clouds play a fundamental role in the Earth’s climate system. Given their importance, understanding and articulating the underlying physics of clouds is critical to making accurate weather and climate predictions. Here we expand the physical understanding of clouds by investigating the interaction between local air circulations associated with individual clouds, allowing us to gain insight into how clouds may interact with themselves and neighboring clouds.  In doing so, we can explain how these cloud-scale motions alter a cloud’s size, intensity, and evolution, thereby providing insight into why clouds achieve certain configurations that we observe in nature. 
%




\section{Introduction}
Cumulus clouds are among the most influential communicators in Earth’s climate system. These clouds alter the vertical stratification of most environments, thereby connecting boundary layer processes to those of the free troposphere, and generating the horizontal and vertical transfer of the Earth’s moisture, heat, and momentum. Cumulus clouds thus modify the spatial and temporal distributions of precipitation and surface temperatures while also influencing scales of motion ranging from global circulations to local storm systems \citep{khvorostyanov_thermodynamics_2014}. The interaction between cumulus clouds and the environment, both at a regional and a global scale, makes modeling them correctly essential for increasing our confidence in climate projections \citep{bony_clouds_2015}.  Yet, despite steady improvements in computational capabilities, cloud/climate interaction remains one of the largest uncertainties in future climate predictions. Thus, improving computational models and developing better insights into Earth’s climate system necessitates novel approaches to how the physics of cumulus clouds is parameterized \citep{arakawa_cumulus_2004}.

The ability of cumulus clouds to communicate across scales stems from their propensity to generate feedback with themselves and the surrounding environment. These feedback mechanisms facilitate the organization of clouds into larger storm systems that exhibit greater meteorological and climatological impacts \citep{bony_clouds_2015}.
Recently, large eddy simulations (LES) have attributed convective organization to cold pool dynamics \citep{schlemmer_formation_2014}, competition for mixed-layer resources \citep{chen_role_2023}, free troposphere resource transport due to cloud advection \citep{holloway_moisture_2009}, and the dry kinematic interaction associated with the flows generated by cumulus clouds \citep{chen_role_2023}. Although LES serves as a useful tool for studying convective organization, its prohibitive computational expense
necessitates alternative approaches to dynamically parameterize sub-grid cloud processes in climate models. Generating simple solutions from conceptual models is a pivotal step toward developing such adaptable parameterizations. Existing conceptual models, and the parameterizations they yield, are limited by their steady-state assumption and\slash{}or an incomplete treatment of cloud-scale horizontal motion.

Existing cumulus cloud conceptual models can be divided into two broad groups: the steady-state plume and the thermal chain. Although the steady-state plume serves as the standard model from which most parameterizations are built, it cannot capture the temporal evolution of individual real-world clouds \citep{scorer_bubble_1953}.  Furthermore, traditional plume models assume homogeneity within regions of ascent and descent, thereby precluding representation of the horizontal heterogeneity that arises from the mixing of environment and cloudy air at a cloud's edge \citep{de_rooy_entrainment_2013}. Although these limitations hinder plume models' applicability to individual clouds, they are generally accepted as useful models for describing a cumulus cloud field.  Fields of cumulus clouds are parameterized by considering an ensemble of plume-like clouds under the assumption that horizontal cumulus eddy transports are negligible compared to the large-scale flow \citep{arakawa_interaction_1974} or by treating the bulk motion of the cloud field as plume-like \citep{tiedtke_comprehensive_1989}.  In either case, the application of the plume model to a cloud field ignores the horizontal interactions between cloud-scale motions.
Thermal chain models use the observational fact that cloud updrafts are comprised of a series of rising buoyant thermals \citep{morrison_thermal_2020,scorer_bubble_1953}. Each of these buoyant thermals can be thought of as a Hill's spherical vortex, wherein motions are poloidal, with positive toroidal vorticity within the sphere and zero toroidal vorticity outside of it.  
Despite the utility of thermals and their representation via Hill's vortices, they are not without their limitations; namely, thermals only describe the near-field motion of an individual buoyant element, not the aggregate circulation of a cumulus cloud.  

To  understand the total cumulus cloud circulation, how it interacts with itself, and with that of neighboring clouds, we propose
a new conceptual model for the total three-dimensional velocity field of a single cumulus cloud, which we refer to as the Kinematic Representation of a Non-rotating Updraft Tori (KRoNUT) model \citep{igel_nontraditional_2020}; where the description ``non-rotating'' emphasizes the absence of rotation in the horizontal direction. Rather than focusing on the individual motions of thermals, the KRoNUT serves as a kinematic representation of the composite flow field generated by a series of thermals that comprise a thermal chain over a cumulus cloud's turnover timescale (on the order of minutes). By focusing on the collective thermal motions over these time scales, the KRoNUT framework describes the local ascent that makes up a cumulus cloud's updraft as well as the non-local aspects of the cloud scale flow such as dynamic detrainment aloft, subsidence outside of the convective updraft, and dynamic entrainment at cloud base, see \citep{de_rooy_entrainment_2013} for definitions.   The parameters of a single KRoNUT describe its maximum vertical velocity, horizontal and vertical extent, which are allowed to change in time.   Using a moment reduction of the Navier-Stokes equations in the presence of a KRoNUT circulation, we derive an analytic framework for the time evolution of the properties of a KRoNUT, under self-advection and the advections of other KRoNUTs which comprise a field of clouds.

To develop this analytic framework, we begin by contextualizing the scope of this work as a series of relaxation experiments under various forcings, Section 2. We then reintroduce the mathematical representation of the KRoNUT model in Section 3. In Section 4, we discuss the moment reduction technique and derive the equations needed to implement the technique. In Section 5, we use the moment reduction technique to generate a system of ordinary differential equations that describe the temporal evolution of the parameters governing a KRoNUTs geometry, intensity, and location. These are the Dynamics of a Non-rotating Updraft Tori (DoNUT) equations. To better understand the behavior associated with the DoNUT equations, we perform a phase portrait analysis and examine multiple test cases in section 5. 
In section 6, we discuss the DoNUT equations in the context of atmospheric inversions and variable turbulent viscosity.
A summary of our findings is listed in Section 7, and we provide concluding remarks in  Section 8. Appendix A explains the notation, which can be easily consulted when reading the paper.

\FloatBarrier
\section{Cloud-Scale Motion Forcings and the Relaxation Experiments}\label{section: Exposition}
\subsection{Assumptions}
The assumptions of the KRoNUT framework are as follows:

\begin{enumerate}
    \setcounter{enumi}{0}
    \item Cumulus clouds are well-separated with compact vorticity distributions that rapidly decay outside of a cloud's updraft, allowing us to split the total vorticity budget into equations for each cloud.  These cloud-specific vorticity equations are thus approximately true in the vicinity of that cloud's convective core.\label{assumption 2}
    \item  A cloud's vorticity, $\vec{\omega}_g$, is in the horizontal direction and remains locally axisymmetric about the z-axis for all time.  Axisymmetry is an invariant in the single KRoNUT case without further assumption, and enforced in the multi KRoNUT case by removing non-azimuthal advection and deformation terms. \label{assumption 3}
    \item Each cloud's velocity field, $\vec{u}_g$, is incompressible $(\nabla\cdot\vec{u}_g=0)$ and thus solenoidal, implying that the velocity field is described entirely by a vector potential, $\vec{\psi}_g$. Future work looks to relax this assumption to the anelastic approximation.  \label{assumption 4}
  \item  The turbulent kinematic viscosity, $\nu$, is a bulk property of the fluid and is constant in the range of $1-10^3 m^2 s^{-1}$.  In reality, $\nu$ likely depends on the individual cloud-scale circulation itself and may vary in space and time. This level of detail, however, exceeds the scope of this work.\label{assumption 1}
\end{enumerate}

\subsection{Forcings and Relaxation}
To better explain the motivation and scope of this paper, we now describe how the forcings acting on an individual cumulus cloud affect its circulation. Taking the curl of the Navier-Stokes equations, the vorticity equation is
\begin{equation}\label{total vorticity}
    \partial_t\vec{\omega}_T=\nabla\times\left(\vec{u}_T\times\vec{\omega}_T\right)+\frac{\nabla\rho\times\nabla p}{\rho^2}+\nabla\times\left(\nu\nabla^2\vec{u}_T\right)
\end{equation}
where $\vec{\omega}_T$ is the total vorticity field associated with the cloud field, $\vec{u}_T$ denotes the total velocity associated with the cloud field, $\rho$ describes the density, $p$, the pressure, and the last term represents the parameterization of turbulent diffusion of vorticity by sub-cloud scale eddies. Using Assumptions \ref{assumption 2} and \ref{assumption 1}, we split the total vorticity budget into the local contributions from each cloud in the cloud field, 
\begin{equation}\label{full near field vorticity}
    \overbrace{\partial_t\vec{\omega}_n}^\text{Term I}=\underbrace{\nabla\times\left(\vec{u}_n\times\vec{\omega}_n\right)}_\text{Term II}+\overbrace{\sum\nabla\times\left(\vec{u}_f\times\vec{\omega}_n\right)}^\text{Term III}+\underbrace{\nu\nabla^2\vec{\omega}_n}_\text{Term IV}+\overbrace{\frac{\nabla\rho\times\nabla p}{\rho^2}}^\text{Term V}
\end{equation}
where a subscript $n$ denotes the near-field cloud in consideration and a subscript $f$ represents the terms associated with the surrounding clouds in the cloud field. Term I describes the near-field cloud's vorticity tendency, Term II and Term III describe advection, deformation, and stretching for the self-interaction of a KRoNUT, and the cross-interaction of the near-field KRoNUT with the other KRoNUTS in the field. Term IV expresses the turbulent diffusion.  Finally, Term V is the baroclinic vorticity generation term. Here, the $\Sigma$ in Term III indicates the summation over all cross-interactions between the near-field cloud vorticity and the velocity field of all of the other clouds. Each cloud in a cloud field is described by its own version of (\ref{full near field vorticity}).

Understanding a cumulus cloud's life cycle from genesis to extinction requires a thorough treatment of all terms in (\ref{full near field vorticity}); however, developing the machinery to appropriately describe the interplay between the dynamics and the thermodynamics of a cloud exceeds the scope of a single paper. Instead, in this paper we conduct a series of relaxation experiments where we assume that all of the terms in (\ref{full near field vorticity}) had been active for some time prior to our integration of the equations, thereby allowing a cloud-scale circulation to develop within a cloud field before we systematically ``turn off'' the forcings associated with baroclinic vorticity generation (Term V) and, in some experiments, the forcing associated with the coupling of a KRoNUT with others (Term III)  or turbulent diffusion (Term IV). We then use the KRoNUT framework in combination with a moment reduction technique to investigate the dynamics associated with a cloud-scale flow's relaxation under the influence of advection/stretching. As a point of emphasis, in every example we consider self-interaction (Term II), and do not consider baroclinic vorticity generation (Term V). 
For simplicity, when conducting experiments involving cross-interactions (Term III), we only considered a cloud field consisting of two clouds; however, the framework can be extended to any number of interacting cloud-scale flows. 

Although the generation of vorticity through moist thermodynamic processes is of paramount importance to the evolution of real-world clouds, considering only how flows evolve through advection provides crucial insights into how a three-dimensional cloud-scale flow interacts both with itself and with its neighbors. For example, the experiments we consider address whether a cloud-scale flow in the absence of baroclinicity can increase its maximum vertical velocity (invigorate itself), and how the presence of viscosity and the velocity field of other cumulus clouds influences the potential invigoration of a cumulus cloud. Although the moment reduction technique used in this paper is done without consideration of the thermodynamics, by introducing this framework, we set the stage for future work in which the thermodynamics and dynamic processes will be actively coupled via the baroclinic term.    

\section{The KRoNUT Model}\label{subsection: KRoNUT Model}

\subsection{Functional Form}
\begin{figure}[h]
 \centerline{\includegraphics[width=33pc]{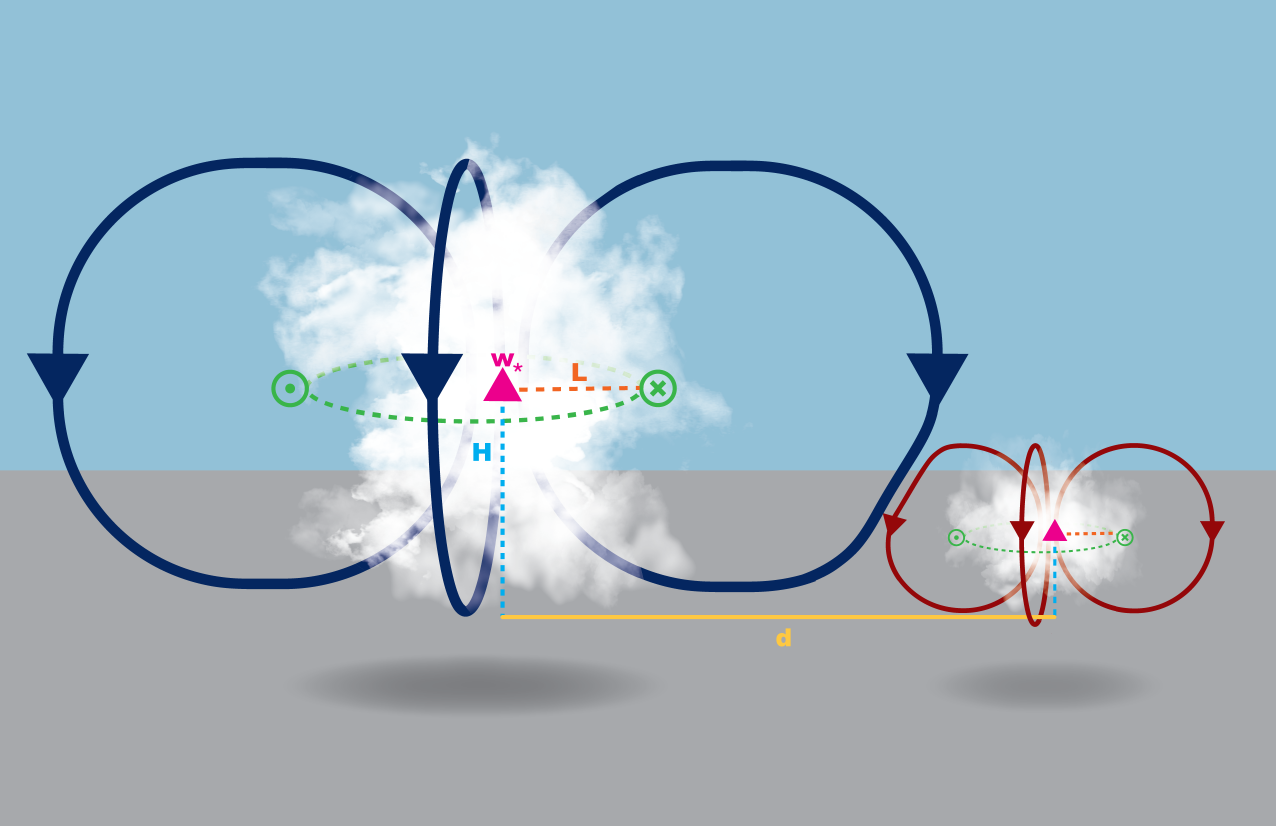}}
  \caption{A simple schematic illustrating KRoNUT parameters and their relationship to coupled cloud-scale motions. Here, a cloud's streamlines are indicated by navy blue or red, the geometric KRoNUT parameters, $L$ and $H$ are denoted by orange and cyan dashed lines, respectively, while a KRoNUT's intensity parameter, $w_*$, is represented by the magenta arrow. The separation between KRoNUTs, $d$, is marked by a solid yellow line, and a dashed green line represents the vorticity line at the $r=L$ and $z=H$.}\label{Schematic}
\end{figure}

\begin{figure}[h]
 \centerline{\includegraphics[width=39pc]{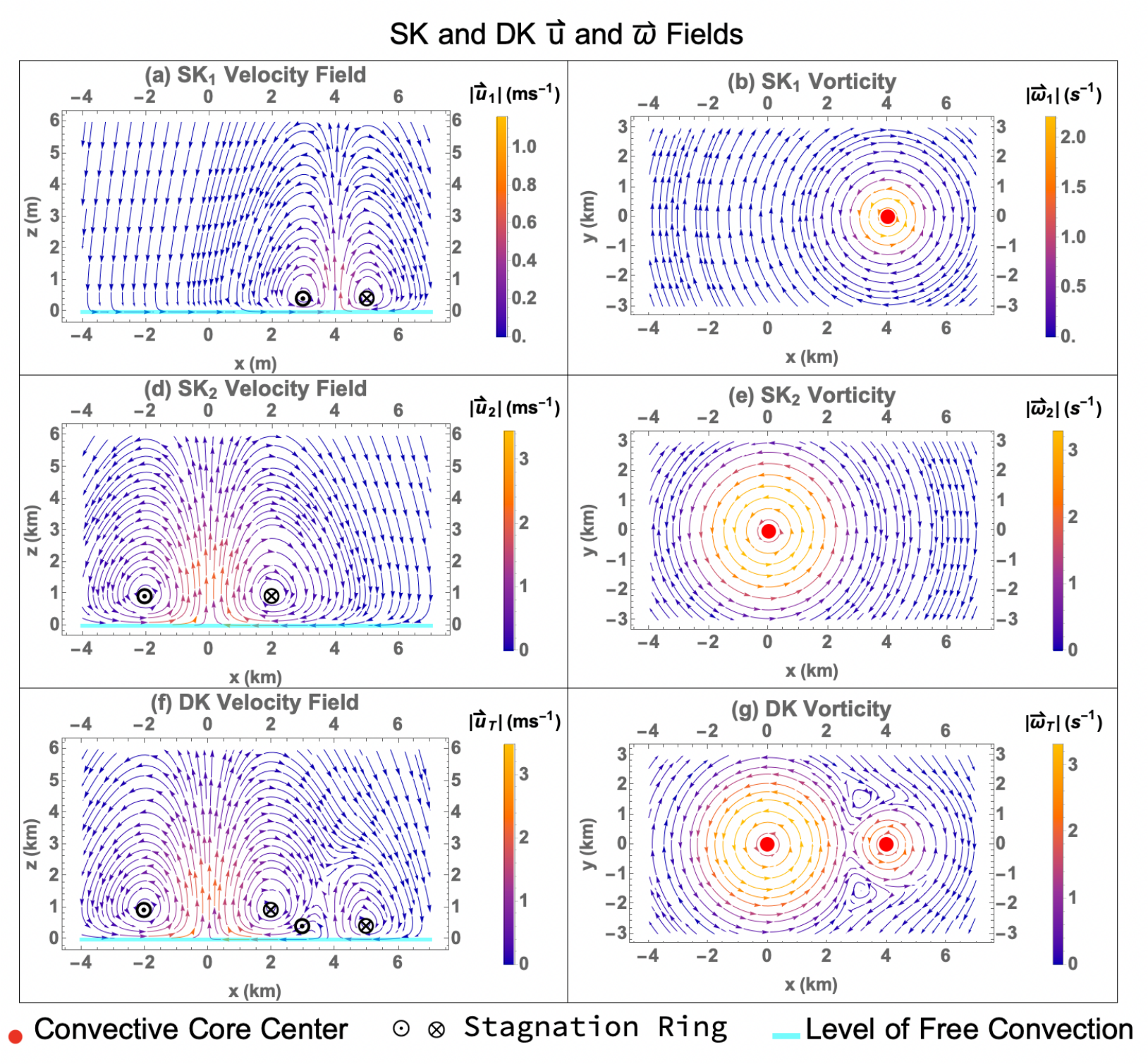}}
  \caption{Example KRoNUT fields.  The left column shows the $y=0$ cross-sections of the streamlines.  The right column shows the $z=H$ cross-sections of the vortex lines. The first two rows are individual KRoNUT circulations, and the bottom row shows the superposition KRoNUTs at a given separation. In all plots, color indicates the magnitude and field line lengths are arbitrary.}\label{SK and DK Fields}
\end{figure}

The KRoNUT model presented here, which has a slightly faster radial decay rate than that considered in \cite{igel_nontraditional_2020}, corresponds to a five-parameter toroidal vector potential, $\vec{\psi}_g$,  
    \begin{equation}
        \vec{\psi}_g =\frac{\text{w}_{*g} z}{2\text{H}_g}e^{1-\frac{z}{\text{H}_g}-\frac{\left(\Tilde{x}-x_g\right)^2+\left(\Tilde{y}-y_g\right)^2}{\text{L}_g^2}}
        \left\{-
        (\tilde{y} - y_g ) \hat{x} + \left(\tilde{x} - x_g\right) \hat{y} \right\}
        \label{general vector potential}
    \end{equation}
which, by Assumption \ref{assumption 4}, yields a poloidal velocity field 
\begin{equation}\label{velocity vector from potential}
    \vec{u}_g=\nabla\times\vec{\psi}_g
\end{equation}
and compact toroidal vorticity field,
\begin{equation}\label{vorticity from vector potential}
    \vec{\omega}_g=\nabla\times\vec{u_g}=-\vec{\nabla}^2\vec{\psi}_g.
\end{equation}
Here $\Tilde{x}$, $\Tilde{y}$, and $\Tilde{z}=z$ are global Cartesian variables and $\hat{x}$ and $\hat{y}$ are the usual Cartesian unit vectors.  While the spatial dependence is evident in (\ref{general vector potential}), the temporal dependence appears implicitly in the parameters $x_g(t)$ and $y_g(t)$ which are the horizontal location of the convective core, $L_g(t)$, the radius of the convective updraft, $H_g(t)$, the height of the maximum vertical velocity in the convective updraft, and $w_{*g}(t)$, the magnitude of the maximum vertical velocity in the convective updraft. For a schematic representation of the KRoNUT parameters and the interaction of two KRoNUTs, see Fig. \ref{Schematic}, and for a graphical representation of $\vec{u}_g$ and $\vec{\omega}_g$ for KRoNUTs $1$ and $2$, see Fig. \ref{SK and DK Fields}.

\begin{figure}[h]
 \centerline{\includegraphics[width=39pc]{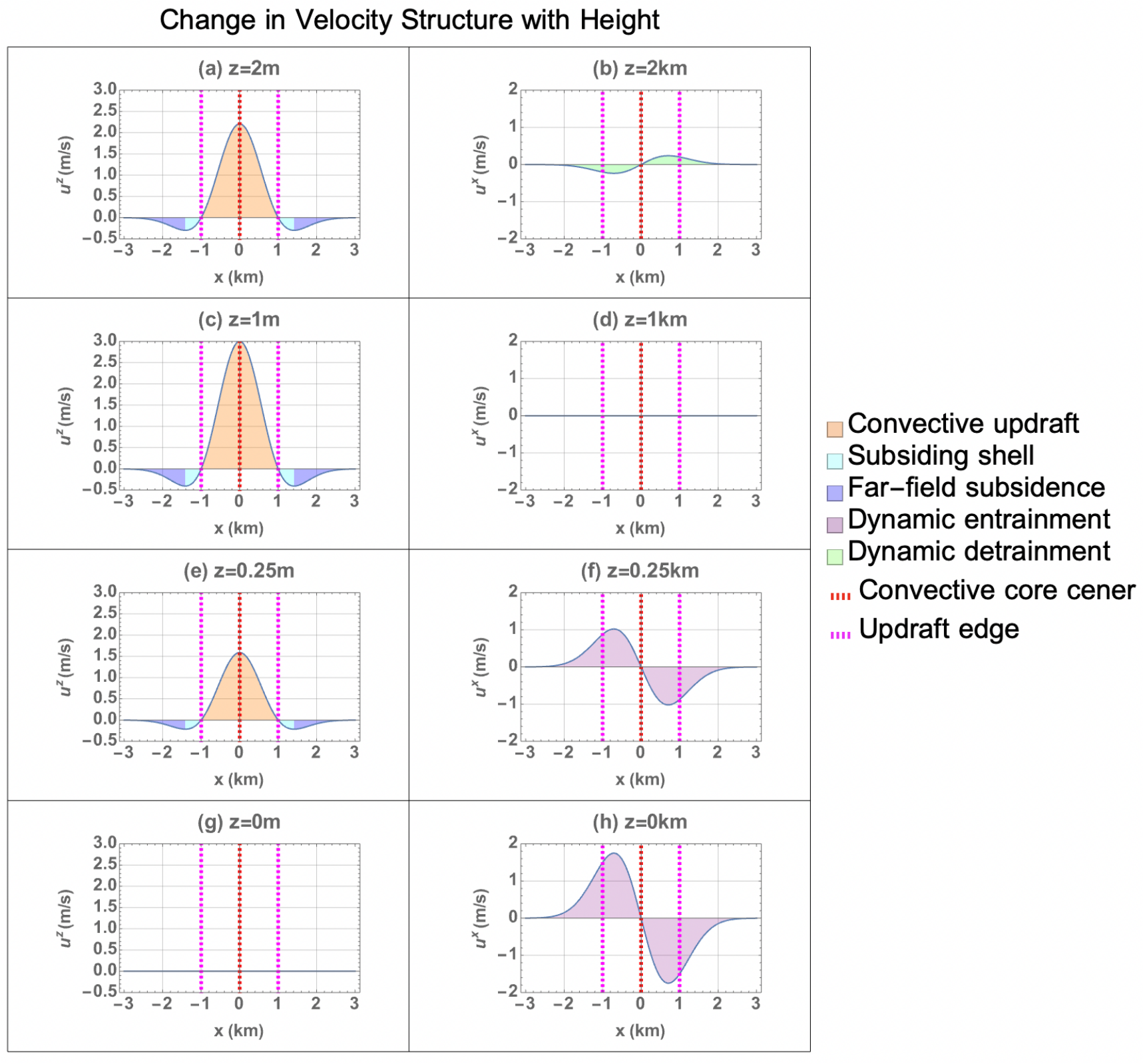}}
  \caption{Horizontal slices of the structural change in a KRoNUT's velocity field at an instant in time for various heights. Here, the left column corresponds to the vertical velocity, and the right column coincides with the horizontal velocity. The initial conditions for this KRoNUT are $w_{*g}=3 m s^{-1}$, $L_g=1000 m$, $H_g=1000m$, $x_g=0m$, and $y_g=0m$.}\label{Velocity Field Cross Sections}
\end{figure}

Despite its relatively simple construction, this KRoNUT model's velocity field emulates the salient features of a pre-existing cumulus cloud-scale flow. Namely, in an actual cloud-scale flow, the buoyancy generated by the latent heating of condensation drives the positive vertical velocity associated with the cloud's updraft region. This positive vertical velocity achieves its maximum value near the cloud's convective core center before decaying radially as mixing with environmental air increases the dilution of buoyancy. Continuing to move radially outward, the vertical velocity reaches zero at the cloud's edge, which by construction occurs at $r=L_g$ in this KRoNUT model, and then transitions to a relatively strong, negative region formed by the latent cooling of evaporation known as the subsiding shell. Beyond the subsiding shell, the negative vertical velocity of the cloud-scale flow asymptotes to zero, constituting its region of far-field subsidence; see the left-hand column of Fig. \ref{Velocity Field Cross Sections} for a visual representation of how a KRoNUT's vertical velocity profiles various height at a given instant in time. The prescribed form of the KRoNUT's vertical velocity profile aligns well with that observed in` \cite{heus_subsiding_2008}.

Linking these regions of ascending and descending air is the horizontal motion associated with cloud-scale dynamic entrainment and detrainment. By construction, this KRoNUT model transitions between dynamic entrainment and detrainment at the height, $z=H_g$, thereby capturing low-level inflow and upper-level outflow. This KRoNUT model's representation of these horizontal motions at various heights for a given instant in time is shown in the right-hand column of Fig. \ref{Velocity Field Cross Sections}. Although a KRoNUT model does not explicitly represent sub-cloud-scale eddies, by representing a cloud's velocity field as a continuously twice-differentiable analytic function, the physics of these processes, namely turbulent entrainment and detrainment, are parameterized by the turbulent diffusion of momentum and thus vorticity.

 By incorporating the essence of these cloud-scale motions into the kinematics of the flow, this KRoNUT model well represents an arbitrary morphology of pre-existing cumulus convection while using a relatively low number of time-dependent parameters to describe its geometry, intensity, and location. Additionally, unlike other conceptual models, a KRoNUT model does not require steady-state assumptions to achieve a closure. As such, as a KRoNUT evolves in time, it can interact with its own flow as well as the KRoNUTs associated with neighboring cloud-scale flows, making it a unique tool to investigate cloud-scale dynamics. 
 
\FloatBarrier
\subsection{Current KRoNUT Limitations}
Despite the versatility of this KRoNUT model, it comes with a few caveats. First and foremost, the KRoNUT does not represent all aspects of convection. Rather, it captures the spatial and temporal average flow generated at the cloud scale. As such, it ignores sub-cloud scale eddies but treats the bulk of their resulting motions.  In its current construction, $z=0$ corresponds to the LFC where, in our model, a no-penetration and a free-slip boundary condition are enforced. Thus, the current KRoNUT model is decoupled from the planetary boundary layer, thereby neglecting processes such as boundary layer inflow into the free troposphere, competition between clouds for mixed-layer resources, and surface forcings. Furthermore, this KRoNUT model lacks the freedom to detach from the boundary layer. Together, these suggest that this KRoNUT model may well represent some aspects of active convection but fails to encompass key features associated with forced or passive convection as defined by \cite{stull_fair-weather_1985}. 
In addition to a decoupling from the boundary layer, the current KRoNUT form lacks a tropopause. While this likely has little effect on shallow clouds, the absence of an up boundary allows deeper cloud types can reach unreasonable heights, as will be shown in Section 5. Finally, while the current KRoNUT represents the three-dimensional flow at the cloud-scale, its prescribed form does not account for convective core tilt or ellipticity. Addressing these limitations, expanding the forcings considered in Section 2, and relaxing the assumptions used in this work are areas of active research by the authors.  

\FloatBarrier

\section{Moment Reduction Technique}
\subsection{An Overview of the Moment Reduction Technique}
To develop a KRoNUT's dynamics, we perform a moment reduction inspired by \cite{morrison_hamiltonian_2009}, \cite{meacham_hamiltonian_1997}, and \cite{melander_moment_1986}. Importantly, this moment reduction must be done with respect to the conserved quantity derived from the vorticity budget equation, that is the circulation density which we denote $\gamma$, (see Subsection \ref{The conservation of circulation density} below).  In doing so, we generate the approximate time rate of change for the parameters that govern the structure and magnitude of a KRoNUT's flow (i.e. its DoNUT equations). Such a reduction is effected by equating the total derivative of a moment of the conserved quantity to the corresponding moment of the local time rate of change associated with that conserved quantity,  
\begin{equation}\label{general equivalence of moments}
    \begin{aligned}
        \dot{M}_\kappa=\frac{d}{dt} \overline{(\kappa \gamma)}&=\overline{\kappa\partial_t(\gamma)}=N_{\kappa}
    \end{aligned}
\end{equation}
where $M$ describes the moment of the conserved quantity, $\gamma$, $N$ represents the moment of the local time rate of change for this conserved quantity, $\kappa$ denotes the weight associated with a particular moment, and an accented dot denotes a total time derivative (i.e. $\frac{dM_k}{dt}=\dot{M_k}$). Here, an overbar indicates the volume integral over the entire extent of the atmosphere for an arbitrary quantity $a$. That is
\begin{equation}\label{Cartesian spatial average}
    \overline{a}
    = \int_0^{2 \pi}
    \int_0^{\infty} \int_0^{\infty} a \; r dr dz d\theta.
\end{equation}

Any moment reduction of non-linear equations suffers from a closure problem, wherein the time tendency of one moment is related to a higher moment. In the DoNUT framework, this problem is obviated by the selection of a spatial structure, in this case $\vec{\psi}_g$, whose parameters are allowed to evolve in time.   Physically, this restricts flow evolution such that a KRoNUT must evolve to a KRoNUT.   In its current form, a single KRoNUT (SK)  only has three parameters (the center does not evolve in time) while a double KRoNUT (DK) has ten parameters (the center of each KRoNUT can evolve in time), as such we consider three or ten moments, respectively.  

We choose the lowest order moments, which have straightforward interpretations, namely the integrated vorticity, $\kappa=r^0$,  the vorticity centroid, $\kappa=\Tilde{x}$ and $\kappa=\Tilde{y}$, and the horizontal and vertical extents of the vorticity distribution, $\kappa=r^2$ and $\kappa=z$, respectively.  In the case of the SK configuration, axisymmetry about the convective center allows for a reduction to three parameters, $w_{*g}$, $L_g$, and $H_g$ which we diagnose by taking moments with weights, $\kappa= r^0, z, r^2$. In the DK case, this symmetry breaks down, but by using the well-separated vorticity Assumption \ref{assumption 2}, we can perform the reduction locally in the vicinity of each KRoNUT's convective core. Five moments suffice to diagnose the local parameters, $w_{*n}, L_n, H_n, x_n$, and $y_n$ for each KRoNUT,
yielding a tenth-order dynamical system for the pair of KRoNUTs. For the DK  configuration, we use the SK weights plus $\Tilde{x}$ and $\Tilde{y}$.

\subsection{The Conservation of Circulation Density}\label{The conservation of circulation density}
To derive the conserved quantity associated with the SK vorticity budget, we begin with the SK vorticity form of the barotropic Euler equations ($\nu = 0$,  $\vec{u}_f=0$, and baroclinic term turned off in (\ref{full near field vorticity})), 
\begin{equation}
    \partial_t \omega^{\theta} + 
    \vec{u}_T \cdot \nabla \omega^{\theta}
    - \frac{\omega^{\theta} u^{r}_T}{r} = 0,
    \label{vorticity_theta}
\end{equation}
which describes the advection and the cyclostrophic generation of azimuthal vorticity. Multiplying (\ref{vorticity_theta}) by the integrating factor $(2 \pi r)^{-1}$ and defining
\begin{equation}\label{vorticity density}
\gamma = \frac{\omega^{\theta}}{2 \pi r}.
\end{equation}
as the circulation density, we find that $\gamma$ satisfies 
the transport equation or, by virtue of the compressibility assumption, the conservation law
\begin{equation}
    \partial_t \gamma + \nabla \cdot
     \left( \gamma \vec{u}_T   \right) = 0.
    \label{vorticity_theta_2}
\end{equation}
It is clear from equation (\ref{vorticity_theta_2}) that the volume integral of 
$\gamma$ is a constant of the motion for isolated poloidal circulations, so taking moments of $\gamma$ preserves the conservation law in the DoNUT equations. 

We now provide an explanation for why $\gamma$ corresponds to the circulation density and its relation to the moments. This involves recasting the vorticity form of the circulation definition, for a SK in terms of a volume integral over the entire atmosphere. Starting with the vorticity form of the circulation 
\begin{equation}
    \begin{split}
        \Gamma&\equiv\oint_\mathcal{C}\vec{u}\cdot\hat{t}dl\\
        &=\iint_\mathcal{S}\vec{\omega}\cdot\hat{n}dS
    \end{split}
\end{equation}
where $\mathcal{C}$ denotes the circuit encompassing the flow, $\hat{t}$ is the unit tangent vector at every point along $\mathcal{C}$, $dl$ is the differential arc length along $\mathcal{C}$, $\mathcal{S}$ is the surface whose boundary is formed by $\mathcal{C}$, and $\hat{n}$ is the normal vector to $\mathcal{S}$.
We choose $\mathcal{C}$ to be the circuit that runs along the base of the cloud, $z=0$, from $r \rightarrow \infty$ to $r=0$, then up along the symmetry axis, $r=0$, from $z=0$ to $z \rightarrow \infty$, then closing along a quarter circle in the $(r,z)$ plane where $\sqrt{r^2 + z^2} \rightarrow \infty$. The circulation thus takes the form
\begin{equation}\label{SK circulation}
    \Gamma=\int_0^\infty\int_0^\infty\omega^\theta drdz.
\end{equation}
To transform the surface integral in (\ref{SK circulation}) to a volume integral over the entire atmosphere, which we denote $\mathcal{V}$, we multiply the integrand in (\ref{SK circulation}) by $\frac{r}{r}$ to incorporate the cylindrical coordinate system Jacobian and by $\frac{2\pi}{2\pi}$ where the $2\pi$ in the numerator is written as an azimuthal integral for an integrand which has no $\theta$ dependence. We thus rewrite (\ref{SK circulation}) as, 
\begin{equation}\label{SK circulation volume}
    \begin{split}
        \Gamma&=\int_0^{2\pi}\int_0^\infty\int_0^\infty\frac{\omega^\theta}{2\pi r} r drdzd\theta\\
        &=\int_\mathcal{V}\gamma dV\\
        &=\bar{\gamma}
    \end{split}
\end{equation}
where we have used (\ref{Cartesian spatial average}) to write the final line in (\ref{SK circulation volume}), thereby highlighting the fact that the circulation of an individual KRoNUT corresponds to the $r^0$ moment of $\gamma$. The above makes it clear that $\gamma$ corresponds to the circulation per infinitesimal unit of volume, elucidating the physical rationale for the nomenclature of circulation density.  
 \FloatBarrier

\subsection{ The Circulation Density Budget Equation}
 We now derive the appropriate equation for $\gamma_n$. This requires manipulating the most general form of the vorticity equation considered in our relaxation experiments (i.e. viscous DK). We begin with the barotropic form of (\ref{full near field vorticity}) for two interacting KRoNUTs,    
\begin{equation}\label{local vorticity budget}
    \overbrace{\partial_t \vec{\omega}_n}^\text{Term I}=-\underbrace{(\vec{u}_n\cdot\nabla)\vec{\omega}_n}_\text{Term IIa}+\overbrace{(\vec{\omega}_n\cdot\nabla)\vec{u}_n}^\text{Term IIb}-\underbrace{(\vec{u}_f\cdot\nabla)\vec{\omega}_n}_\text{Term IIIa}+\overbrace{(\vec{\omega}_n\cdot\nabla)\vec{u}_f}^\text{Term IIIb}+\underbrace{\nu\vec{\nabla}^2(\vec{\omega}_n)}_\text{Term IV}
\end{equation}
where we have expanded Terms II and III in (\ref{full near field vorticity}) and used the incompressibility condition to remove the divergence terms. The expanded terms on the right-hand side of equation (\ref{local vorticity budget}) describe vorticity advection, Term IIa, and deformation, Term IIb, by a KRoNUT's own velocity field, and vorticity advection, Term IIIa, deformation, Term IIIb, by the neighboring KRoNUT's velocity field.  
 
We now adopt a local cylindrical coordinate system with its origin located at $(x_n,y_n,0)$. In doing so, the SK model is constructed so that there is no $\theta$ dependence in the near field. Thus, the kinematics of vector quantities described in (\ref{local vorticity budget}) have the corresponding components, see Table \ref{table: Vector & Components}, when described in the near field coordinate system.

\begin{table}[h]
\begin{center}
\caption{Components of vector quantities when described in a local cylindrical coordinate system with its origin centered at $(x_n,y_n,)$.}
\label{table: Vector & Components}
\begin{tabular}{|c|c|} 
  \hline
        Vector & Components \\\hline $\vec{\omega}_n(r,z)$ & $\omega_n^{\theta}\hat{\theta}$\\\hline $\vec{u}_n(r,z)$ & $u_n^{r}\hat{r}+u_n^{z}\hat{k}$\\\hline  $\vec{u}_f(r,\theta,z)$ & $u_f^{r}\hat{r}+ u_f^{\theta}\hat{\theta}+  u_f^{z}\hat{k}$\\\hline
\end{tabular}
\end{center}
\end{table}

Using (\ref{vorticity density}) and Table \ref{table: Vector & Components}, we now substitute $\vec{\omega}_n=2\pi r\gamma_n\hat{\theta}$ into (\ref{local vorticity budget}) and derive the local time rate of change of $\gamma_n$.  Recalling that $\gamma_n$ corresponds to the azimuthal component $\vec{\omega}$, this implies taking the dot product $\hat{\theta}\cdot\partial_t\vec{\omega}$.  In doing so, we symmetrically ignore the radial and vertical cross-advection and deformation terms associated with interacting KRoNUTs. Although disregarding these terms, which are not shown in our derivation, implies that the resulting DoNUT equations fail to capture the full physics of cross-advection and deformation, under Assumption \ref{assumption 3}, the errors induced by these terms are small for well-separated KRoNUTs. With this in mind, we begin by noting that a factor of $2\pi$ cancels on either side of (\ref{local vorticity budget}), Term I takes the form,
\begin{equation}\label{term I}
    \partial_t\left(
    r\gamma_n\right)= r\partial_t\gamma_n
\end{equation}
while the implementation of the product rule allows us to reduce term IV to 
\begin{equation}
\hat{\theta} \cdot   \nu\nabla^2\left(r\gamma_n \hat{\theta}\right)=\nu\left(3\partial_r\gamma_n+r\partial_{rr}\gamma_n+r\partial_{zz}\gamma_n\right).
\end{equation}
To diagnose the remaining advection and deformation terms, it is useful to express the azimuthal component of the advective derivative in cylindrical coordinates,
\begin{equation}\label{convective operator}
\hat{\theta}\cdot\left(\vec{b}\cdot\nabla\right)\vec{c}=\left(b^r\partial_r c^\theta+\frac{b^\theta}{r}\partial_\theta c^\theta+b^z\partial_z c^\theta+\frac{b^\theta c^r}{r}\right)\\
\end{equation}
where  $\vec{b} = b^r \hat{r} + b^{\theta} \hat{\theta} + b^z \hat{z}$ and similarly $\vec{c}$.  Using (\ref{convective operator}) to express Term IIa 
we find 
\begin{equation}\label{term IV}
\hat{\theta}\cdot\left(\vec{u}_n\cdot\nabla\right)\left(
             r\gamma_n\hat{\theta}\right)
             =\gamma_n u_n^r +r u_n^r  \partial_r\gamma_n+r u_n^z  \partial_z\gamma_n
\end{equation}
In the same manner, Term IIb becomes
\begin{equation}\label{term III}
\hat{\theta}\cdot\left(\left(r\gamma_n\hat{\theta}\right)\cdot\nabla\right)\vec{u}_n
             =\gamma_n u_n^r
\end{equation}
Notice that the advection and deformation of a KRoNUT by its own circulations leaves the circulation density, and thus vorticity, of a single KRoNUT purely in the $\hat{\theta}$ direction. For the cross-interaction terms, Term IIIa  
assumes the form 
\begin{equation}\label{term IV}
\hat{\theta}\cdot\left(\vec{u}_f\cdot\nabla\right)\left(r\gamma_n\hat{\theta}\right)
             =\gamma_n u_f^r+ r u_f^r\partial_r\gamma_n + u_f^\theta \partial_\theta \gamma_n  + r u_f^z\partial_z\gamma_n
\end{equation}
while  term IIIb becomes
\begin{equation}\label{term III}
            \hat{\theta}\cdot\left(\left( r\gamma_n\hat{\theta}\right)\cdot\nabla\right)\vec{u}_f
            =\gamma_n \left(\partial_\theta u_f^\theta + u_f^r\right)
\end{equation}
Making the appropriate cancellations (\ref{local vorticity budget}) takes the form 
\begin{equation}\label{conserved quantity budget}
    \overbrace{\partial_t\gamma_n}^\text{\clap{Term I: $\gamma_n$ tendency}}=-\underbrace{u_n^r\partial_r\gamma_n -  u_n^z\partial_z\gamma_n}_\text{\clap{Term IIa: Adv. of $\gamma_n$ by $\vec{u}_n$}}+\overbrace{u_f^r\partial_r\gamma_n + \frac{u_f^\theta}{r} \partial_\theta \gamma_n  + u_f^z\partial_z\gamma_n}^\text{\clap{Term IIIa: Adv. of $\gamma_n$ by $\vec{u}_f$}}+\underbrace{\frac{\gamma_n}{r}\partial_\theta u_f^\theta}_\text{\clap{Term IIIb: Def. of $\gamma_n$ by $\vec{u}_f$}}+\overbrace{\nu\left( \frac{3}{r} \partial_r \gamma_n + \partial_{rr}\gamma_n + \partial_{zz}\gamma_n\right)}^\text{\clap{Term IV: Dif. of $\gamma_n$}}
\end{equation}
where Term I denotes the spatial description of the $\gamma_n$ tendency, Term II corresponds to the effects of turbulent diffusion on $\gamma_n$, Term III accounts for self-advection of $\gamma_n$, Term IV indicates cross-deformation of $\gamma_n$, and Term V treats cross-advection of $\gamma_n$. 

\FloatBarrier
\section{Generating the Dynamical systems}\label{section: dynamical systems}
\subsection{Time Derivative of the Circulation Density Moments}

The time derivative of the circulation density moments corresponds to the left-hand side of (\ref{general equivalence of moments}). Since we are treating the near field, the subscripts on (\ref{general vector potential}) go from $g$ to $n$. To exploit the cylindrical symmetry in this vector potential, we start by performing a transform of variables,
\begin{subequations}\label{p transform of variables}
    \begin{align}
        x&=\Tilde{x}-x_n\\
        y&=\Tilde{y}-y_n.
    \end{align}
\end{subequations}
In turn, the horizontal center of KRoNUT$_n$'s convective core, $(x_n,y_n)$, corresponds to the origin. We then transform the coordinates from a Cartesian basis to a cylindrical one with its azimuthal angle about the z-axis. The vector potential then takes the form,
\begin{equation}\label{n vector potential in cylindrical}
\begin{aligned}
    \Vec{\psi}_n(r, z; w_{*n}, L_n, H_n)&=
    \psi_n^{\theta}\hat{\theta}\\
    &=\frac{\text{w}_{*n} r z}{2\text{H}_n}
    e^{1-\frac{z}{\text{H}_n}-\frac{ r^2}{\text{L}_n^2}}\hat{\theta}
\end{aligned}
\end{equation}
where $r=\left(x^2+y^2\right)^\frac{1}{2}$.  Applying (\ref{vorticity from vector potential}) and (\ref{vorticity density}), $\gamma_n$ takes the form, 
\begin{equation}\label{n circulation density}
    \begin{aligned}
       \gamma_n
       &=\frac{w_{*n} e^{1-\frac{z}{\text{H}_n}-\frac{r^2}{\text{L}_n^2}} \left(\text{H}_n^2 \left(8 \text{L}_n^2 z-4 z r^2\right)+2 \text{H}_n \text{L}_n^4-\text{L}_n^4 z\right)}{4 \pi  \text{H}_n^3 \text{L}_n^4}. 
    \end{aligned}
\end{equation}
Substituting (\ref{n circulation density}) into (\ref{Cartesian spatial average}) with the appropriate weights, we arrive at the following moments,
\begin{subequations}\label{circulation density moments}
    \begin{align}
        M_{r^0}&= e \text{w}_{n*}\text{H}_n\left(\frac{\text{L}_n^2}{4 \text{H}_n^2}+1\right)=\Gamma_n
        \label{vorticity moment r^0}\\ 
        M_{\Tilde{x}^1} &= x_n\left(e \text{w}_{n*}\text{H}_n\left(\frac{\text{L}_n^2}{4 \text{H}_n^2}+1\right)\right)=x_n\Gamma_n\\
        M_{\Tilde{y}^1} &=y_n\left(e \text{w}_{n*}\text{H}_n\left(\frac{\text{L}_n^2}{4 \text{H}_n^2}+1\right)\right)=y_n\Gamma_n\label{vorticity moment y}\\
        M_{z^1} &= 2 e \text{H}_n^2 w_{*n} \label{vorticity moment z}\\
        M_{r^2} &= \frac{e \text{L}_n^4 w_{*n}}{4 \text{H}_n}\label{vorticity moment r^2}.
    \end{align}
\end{subequations}
For the SK configuration, we use (\ref{vorticity moment r^0}), (\ref{vorticity moment z}), and (\ref{vorticity moment r^2}), whereas in the DK case, we implement all of (\ref{circulation density moments}) in the moment reduction technique.
\FloatBarrier
\subsection{Single KRoNUT Dynamics}
We now look to treat the appropriate SK terms in the right-hand side of (\ref{general equivalence of moments}). Using the form of the vector potential given by (\ref{n vector potential in cylindrical}), the velocity takes the form, 
\begin{subequations}
    \begin{align}
        u^{r}_n&=-\frac{\text{w}_{*n} r (\text{H}_n-z) e^{1-\frac{z}{\text{H}_n}-\frac{r^2}{\text{L}_n^2}}}{2 \text{H}_n^2}\\
        u^{\theta}_n&=0\\
        u^z_n&=\frac{\text{w}_{*n} z \left(\text{L}_n^2-r^2\right) e^{1-\frac{z}{\text{H}_n}-\frac{r^2}{\text{L}_n^2}}}{\text{H}_n \text{L}_n^2}.
    \end{align}
\end{subequations}
Using the above velocity field and circulation density computed in (\ref{n circulation density}), we take the appropriate moments of (\ref{conserved quantity budget}) for the diffusive SK, 
\begin{subequations}
    \begin{align}
        N_{r^{0} }&=-\frac{e \pi  \nu \text{w}_{*n} \left(32 \text{H}_n^4+8 \text{H}_n^2 \text{L}_n^2-3 \text{L}_n^4\right)}{2 \text{H}_n^3 \text{L}_n^2}\label{diffusive SK budget moment r0}\\
        N_{z^1 }&=-\frac{e^2 w_{*n}^2 \left(8 \text{H}_n^2+\text{L}_n^2\right)}{64 \text{H}_n}+\frac{e \nu w_{*n} \left(\text{L}_n^4-32 \text{H}_n^4\right)}{2 \text{H}_n^2 \text{L}_n^2}\label{diffusive SK budget moment z1}\\
        N_{r^2}&=-\frac{e^2 \text{L}_n^4 w_{*n}^2}{32 \text{H}_n^2}+\frac{3 e \text{L}_n^4 \nu w_{*n}}{4 \text{H}_n^3}\label{diffusive SK budget moment r2}
    \end{align}
\end{subequations}
where the first term on the right-hand side owes its origin to self-advection and the second term results from diffusion except in the case of $D_{r^0}$ which only experiences contributions from diffusion. Equating the time derivatives of (\ref{vorticity moment r^0}), 
 (\ref{vorticity moment z}), and (\ref{vorticity moment r^2}) to (\ref{diffusive SK budget moment r0}), (\ref{diffusive SK budget moment z1}), (\ref{diffusive SK budget moment r2}) respectively, we solve for the parameter tendency equations, thereby generating the dynamical system associated with a single, diffusive cloud-scale circulation,

\begin{subequations}\label{SK Adv Dif Tend}
    \begin{align}
        \dot{w}_{*n}&=-\frac{e w_{*n}^2 \left(-32 \text{H}_n^2 \text{L}_n^2+64 \text{H}_n^4-\text{L}_n^4\right)}{128 \text{H}_n^3 \left(8 \text{H}_n^2+3 \text{L}_n^2\right)}-\frac{\nu w_{*n} \left(160 \text{H}_n^4 \text{L}_n^2-16 \text{H}_n^2 \text{L}_n^4+256 \text{H}_n^6-\text{L}_n^6\right)}{4 \text{H}_n^4 \text{L}_n^2 \left(8 \text{H}_n^2+3 \text{L}_n^2\right)}
        \label{wdot Adv-Dif SK}\\
        \dot{L}_n&=-\frac{5 e  w_{*n}\text{L}_n^3}{32 \text{H}_n \left(8 \text{H}_n^2+3 \text{L}_n^2\right)}+\frac{\nu \left(18 \text{H}_n^2 \text{L}_n^2+16 \text{H}_n^4+\text{L}_n^4\right)}{\text{L}_n \text{H}_n^2 \left(8 \text{H}_n^2+3 \text{L}_n^2\right)}\label{Ldot Adv-Dif SK}\\
        \dot{H}_n&=\frac{e w_{*n} \left(64 \text{H}_n^4+\text{L}_n^4\right)}{128 \text{H}_n^2 \left(8 \text{H}_n^2+3 \text{L}_n^2\right)}+\frac{\nu \left(-4 \text{H}_n^2 \text{L}_n^2+32 \text{H}_n^4+\text{L}_n^4\right)}{4 \text{H}_n^3 \left(8 \text{H}_n^2+3 \text{L}_n^2\right)}\label{Hdot Adv-Dif SK}.
    \end{align}
\end{subequations}
These tendency equations describe the evolution of the flow in terms of the original SK parameters.  Each tendency equation (and all subsequent tendency equations) is arranged so that the role of viscosity is apparent.  In the limit that viscosity is negligible, the SK circulation is conserved and each tendency reduces to just its first term in the right-hand side of (\ref{SK Adv Dif Tend}), which accounts for the effect of self-advection.  To further elucidate the physics associated with the SK, we recast these equations in terms of two dimensionless quantities, $\alpha_g$ and $R_g$, and a dimensional quantity, $\tau_g$,
\begin{subequations}\label{dimensionless quantities}
    \begin{align}
        \alpha_g&\equiv\frac{L_g}{H_g}\label{aspect ratio}\\
        R_g&\equiv \frac{w_{*g}H_g e}{\nu}\label{Reynolds number}\\
        \tau_g&\equiv\frac{H_g}{w_{*g}}\label{turnover time}
    \end{align}
\end{subequations}
where  $\alpha_g$ represents the aspect ratio of geometric parameters, $R_g$ denotes a turbulent Reynolds number for the system, and $\tau_g$ corresponds to the circulation turnover time. We refer to the quantities like these, which arise from explicit parameters in the KRoNUT vector potential, (\ref{general vector potential}), as the implicit parameters of the system. Changing our subscripts from $g$ to $n$, we rearrange (\ref{SK Adv Dif Tend}) such that they are described in terms of scaling factors and the quantities of (\ref{dimensionless quantities}),
\begin{subequations}\label{parameter alpha Reynolds number tendency}
    \begin{align}
        \dot{w}_{*n}&=\frac{e}{\left(3\alpha_n^2+8\right)}\frac{w_{*n}}{\tau_n}\left[\left(\frac{\alpha_n^4}{128}+\frac{\alpha_n}{4}-\frac{1}{2}\right)+\frac{1}{R_n}\left(\frac{\alpha_n^4}{4}+4\alpha_n^2-40-64\alpha_n^{-2}\right)\right]\label{w tendency}\\
        \dot{L}_n&=\frac{e}{\left(3\alpha_n^2+8\right)}w_{*n}\left[\left(\frac{-5\alpha_n^3}{32}\right)+\frac{1}{R_n}\left(\alpha_n^3+18\alpha_n+16\alpha_n^{-1}\right)\right]\label{L tendency}\\
        \dot{H}_n&=\frac{e}{\left(3\alpha_n^2+8\right)}w_{*n}\left[\left(\frac{\alpha_n^4}{128}+\frac{1}{2}\right)+\frac{1}{R_n}\left(\frac{\alpha_n^4}{4}-\alpha_n^2+8\right)\right]\label{H tendency}.
    \end{align}
\end{subequations}

Examining (\ref{parameter alpha Reynolds number tendency}), first we note a common scaling factor, $\frac{e}{3\alpha^2+8}$, which slows any tendency as $\alpha_n$ increases. Second, taking the inviscid limit of (\ref{parameter alpha Reynolds number tendency}), $\lim_{R_n\to\infty}$, we see $\dot{L}_n$ is negative definite, and $\dot{H}_n$ is positive definite, implying that $\dot{\alpha}_n<0$ for all clouds regardless of initial morphology.  By contrast, $\dot{w}_{n*}>0$ for $\alpha_n\gtrapprox 1.724$ and $\dot{w}_{n*}<0$ for $\alpha_n\lessapprox 1.724$, indicating that high aspect ratio clouds invigorate their vertical velocity due solely to self-advection. This invigoration cannot persist, though, since $\dot{\alpha}_n<0$. Thus, as time progresses, clouds tend toward $\alpha_n<1.724$ and enervation. In the low aspect ratio cloud limit $\alpha_n\rightarrow 0 $, $(\dot{w}_{*n},\dot{L}_n,\dot{H}_n)=\frac{ew_{*n}}{16}(-\frac{1}{\tau},0,1)$.  This particular limit illustrates the development of ``pencil-like'' clouds which grow vertically due to advection at the expense of weakening their velocity.  Finally, we emphasize that $\dot{H}_n$ is not simply $w_{*n}$ as might be assumed. Relaxing the inviscid limit, it can be seen in (\ref{w tendency}) that narrow clouds with low Reynolds number, have a strong tendency toward enervation. As we shall see in the next subsection, this leads to the rapid dissipation of a cloud. 

Once again, employing our implicit KRoNUT parameters, (\ref{dimensionless quantities}) helps to elucidate the effects of diffusion and the overall physics of the system.  The tendency equations for these are
\begin{subequations}\label{dimensionless tendencies}
    \begin{align}
        \dot{\alpha}_n&=\frac{e}{\left(3\alpha_n^2+8\right)\tau_n}\left[-\left(\frac{\alpha_n^5}{128}+\frac{5}{64}\alpha_n^3+\frac{1}{2}\alpha_n\right)+\frac{1}{R_n}\left(-\frac{1}{4}\alpha_n^5+2\alpha_n^3+10\alpha_n+16\alpha_n^{-1}\right)\right]\label{alpha dot}\\
        \frac{\dot{R}_n}{R_n}&=\frac{1}{\left(3\alpha_n^2+8\right)\tau_n}
      \left[  \left(\alpha_n^4+\frac{1}{4}\alpha_n^2\right)
       +\frac{1}{R_n}\left(\frac{1}{2}\alpha_n^4+3\alpha_n^2-32-64\alpha_n^{-2}\right)\right]\label{R dot}\\
        \dot{\tau}_n&=\frac{e}{\left(3\alpha_n^2+8\right)}\left[\left(-\frac{\alpha_n^2}{4}+1\right)+\frac{1}{R_n}\left(-5\alpha_n^2+48+64\alpha_n^{-2}\right)\right]\label{tau dot}
    \end{align}
\end{subequations}
where (\ref{dimensionless tendencies}) constitute a unique transform of (\ref{parameter alpha Reynolds number tendency}). While these parameter tendencies are less physically intuitive, analyzing them leads to a more fruitful understanding of the original parameters than that offered by the limited discussion of the previous paragraph.

Applying a phase plane analysis to (\ref{parameter alpha Reynolds number tendency}), we observe no fixed points in the system. Solving for the null surfaces, however, we find that the $\alpha_n$ and $R_n$ null surfaces intersect along,
\begin{equation}\label{SK fixed line}
        \left(\alpha_n^*, R_n^*, \tau_n\right)=\left(2          \sqrt{\frac{1}{3} \left(\sqrt{7}+1\right)},\frac{8}{27} \left(17 \sqrt{7}-5\right),\tau_n\right)\approx\left(2.205, 11.845, \tau_n\right).
\end{equation}
This intersection implies the presence of a fixed line that passes through $\tau_n$-dimension. Performing a linearization of the system along the fixed line, we compute the Jacobian matrix in the $\alpha_n$-$R_n$ plane,

\begin{equation}\label{Jacobian of the fixed line}
    J=\left[\begin{matrix}
        \frac{\partial\dot{\alpha}_n}{\partial\alpha_n}&\frac{\partial\dot{\alpha}_n}{\partial R_p}\\\frac{\partial\dot{R}_n}{\partial\alpha_n}&\frac{\partial\dot{R}_n}{\partial R_n}
    \end{matrix}\right]_{FL}\approx\frac{1}{\tau_n}\left[\begin{matrix}
         -0.381& -0.032 \\
        2.993 & 0.07 \\
        \end{matrix}\right].
\end{equation}
Given that $\tau_n$ scales out of the matrix, we may set it equal to unity and compute the trace, $Tr$, determinant, $Det$ and discriminant, $Disc$ associated with (\ref{Jacobian of the fixed line}). This yields, $Tr=-0.311$, $Det=0.0701$ and $Disc=-0.1837$. Given the sign of these terms, the fixed line corresponds to a stable helical spiral \citep{strogatz_nonlinear_2000}.
Substituting the $\alpha_n$ and $R_n$ associated with the fixed line into (\ref{tau dot}) we find, \begin{equation}
    \dot{\tau}_n=\frac{\left(29 \sqrt{7}+64\right) e}{8 \left(25 \sqrt{7}+71\right)}\approx 0.349.
\end{equation}
Positive $\dot{\tau}_n$ indicates that the trajectory of an arbitrary KRoNUT spirals into a particular value of $\alpha_n$ and $R_n$ while increasing its circulation turnover time.

Interestingly, this fixed line carries a special physical meaning. Recalling that the circulation corresponds to the $r^0$ moment of $\gamma_n$, we rewrite (\ref{vorticity moment r^0}) using (\ref{dimensionless quantities}) so that, 
\begin{equation}\label{alpha Reynolds number circulation}
    \Gamma_n=\nu R_n\left(\frac{\alpha_n^2}{4}+1\right).
\end{equation}
Taking the time derivative of (\ref{alpha Reynolds number circulation}) and recalling the fixed line corresponds to the case where $\dot{\alpha}_n=0$ and $\dot{R}_n=0$, then,
\begin{equation}\label{SK Gamma dot}
    \dot{\Gamma}_n=\nu\left[\dot{R}_n\left(\frac{\alpha_n^2}{4}+1\right)+\frac{\dot{\alpha}_n\alpha_n R_n}{2}\right]=0.
\end{equation}
Thus, the circulation for clouds under the influence of self-advection and turbulent diffusion of circulation density tends toward a steady state that depends on $\nu$. 
Substituting  (\ref{SK fixed line}) into (\ref{alpha Reynolds number circulation}), it becomes apparent that, 
\begin{equation}\label{SK preferred circulation}
    \Gamma_n^*=\frac{8 e \nu}{9} \left(7 \sqrt{7}+11\right)\approx 26.24\nu
\end{equation}
where $\Gamma_n^*$ corresponds to the fixed circulation that is set by the diffusion coefficient. Given the stable spiral nature associated with the SK fixed line, we refer to this circulation as the stable fixed circulation.

The presence of a fixed circulation represents an important result of this paper.  It implies that in the absence of the baroclinic term, the dynamics of an isolated cloud evolve toward a steady state circulation, not a steady state mass flux, which is often used as a closure for plume models. To illustrate this fact, we compute the integrated convective core mass flux through the height of the maximum vertical velocity, which we denote as $\sigma_n$,
\begin{equation}\label{convective core mass flux}
    \begin{split}
        \sigma_n&\equiv\int_0^{L_n}\int_0^{2\pi}u^z_n(r,\theta,H_n)rdrd\theta\\&=\frac{\pi w_{*n} L_n^2}{e}.
    \end{split}
\end{equation}
Taking a time derivative of (\ref{convective core mass flux}), dividing both sides by $\sigma_n$, and substituting in the appropriate implicit parameters we express the logarithmic convective core mass flux  tendency as, 
\begin{equation}\label{mass flux tendency}
    \frac{\dot{\sigma}_n}{\sigma_n}=\frac{e}{(3\alpha_n^2+8)\tau_n}\left[\left(\frac{\alpha_n^4}{128}-\frac{1}{16}\alpha_n^2-\frac{1}{2}\right)+\frac{1}{R_n}\left(\frac{1}{4}\alpha_n^4+6\alpha_n^2-4-32\alpha_n^{-2}\right)\right]
\end{equation}
and substituting (\ref{SK fixed line}) into (\ref{mass flux tendency}), we find, 
\begin{equation}
        \frac{\dot{\sigma}_n}{\sigma_n}=\frac{e\left(51\sqrt{7}-59\right)}{1184\tau_n}\approx \frac{0.174}{\tau_n}.
\end{equation}
This demonstrates that the fixed line and associated steady-state circulation do not coincide with a steady-state mass flux, thus, there is no steady-state mass flux in the system. Additionally, as cloudy circulations asymptote to their respective stable fixed circulations, their mass fluxes increase with time; however, since $\dot{\tau}_n$ is also positive along the fixed line, the rate of mass flux increase also becomes weaker in this infinite time limit.  This suggests that the nature of cloud-scale flows is better understood via the associated circulation and its corresponding fixed line as opposed to the convective core mass flux. 

\subsection{Single KRoNUT Test Cases}
\begin{table}[h]
\caption{Types of cumulus clouds loosely adapted from \cite{khvorostyanov_thermodynamics_2014} Table 2.1 and \cite{houze_cloud_2014} Table 1.1.  Overlap between cloud types is meant to allow for transitions between cloud types.}\label{Cloud Type}
\begin{center}
\begin{tabular}{|c|c|c|c|c|} 
  \hline
  Cloud Type: & $w_*$(m/s) & $L$(m) & $H$(m) & Lifetime (minutes)\\\hline 
  Shallow (Sh):& 1-3 & 500-2000 & 500-2000 & 10-40 \\\hline
  Mid (Mi):& 3-10 & 1500-4500 & 1500-4500 & 20-45\\\hline
  Deep (De):& 5-30 & 4000-8000 & 4000-8000 & 45\\\hline
\end{tabular}
\end{center}
\end{table}

To contextualize $\Gamma_n^*$ and its application to real clouds, we broadly examine particular types of isolated cumulus clouds translated into the $\alpha_n$-$R_n$ phase space. To aid in this investigation, we split cumulus clouds into three groups of characteristic size and intensity, shallow (Sh), mid-level (Mi), and Deep (De) (see Table $\ref{Cloud Type}$). These values are then used to guide the initialization of particular test cases of interest. Importantly, in Table $\ref{Cloud Type}$, $H$, the height of maximum vertical velocity, is not synonymous with the depth of the cloud top. However, it is assumed reasonable to treat $H$ as half of the cloud top height. For simplicity, we have also assumed an $L$, the horizontal updraft radius, with the same range as $H$ for a given cloud type. Considering values of $\nu$ ranging from 1-1000$m^2 s^{-1}$, we use Table \ref{Cloud Type} to define a series of test KRoNUTs, (see Table \ref{Cloud Geometry/Intensity}) for notation. 
\begin{table}[h]
\caption{KRoNUT test case terminology for intensity, viscosity, and geometry}\label{Cloud Geometry/Intensity}
\begin{center}
\begin{tabular}{|c|c|c|} 
  \hline
  Cloud Intensity:& Turbulent Kinematic Viscosity:& Cloud Geometry:\\\hline 
  Strong (S): $ w_{*}\rightarrow High$& Viscous (V): $\nu\rightarrow High$& Mound: $\alpha\rightarrow High$\\\hline
  Weak (W): $ w_{*}\rightarrow Low$& Inviscid (I): $\nu\rightarrow Low$& Dome: $\alpha\approx 1$\\\hline
  & & Tower: $\alpha\rightarrow Low$\\\hline
\end{tabular}
\end{center}
\end{table}
To keep the number of test KRoNUTs manageable, we analyze cases that span the corners of the  $\alpha_n$-$R_n$ phase space for each cloud type defined in Table \ref{Cloud Type}. These test cases are outlined in Tables \ref{Shallow test cloud table}, \ref{Mid test cloud table}, \ref{Deep cloud type table}.

\begin{figure}[h]
 \centerline{\includegraphics[width=27pc]{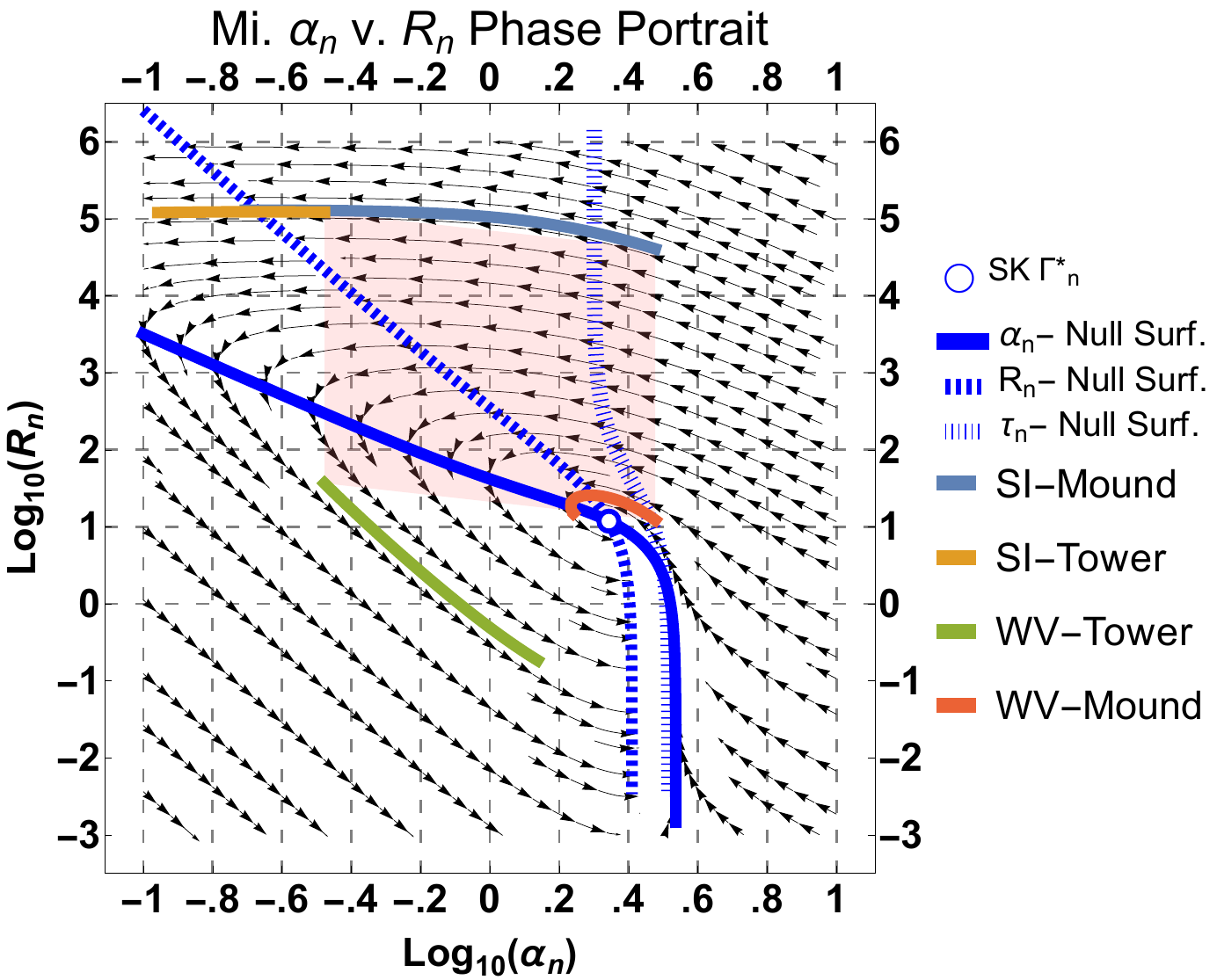}}
  \caption{Three-hour trajectory evolutions of the Mi. cloud-type test cases projected onto the $\alpha_n$-$R_n$ subspace of the phase space. Here the pink-shaded quadrilateral corresponds to the $\alpha_n$-$R_n$ region of the phase space occupied by Mi. convection.}\label{Md SK Phase Portrait}
\end{figure}

\begin{figure}[h]
 \centerline{\includegraphics[width=33pc]{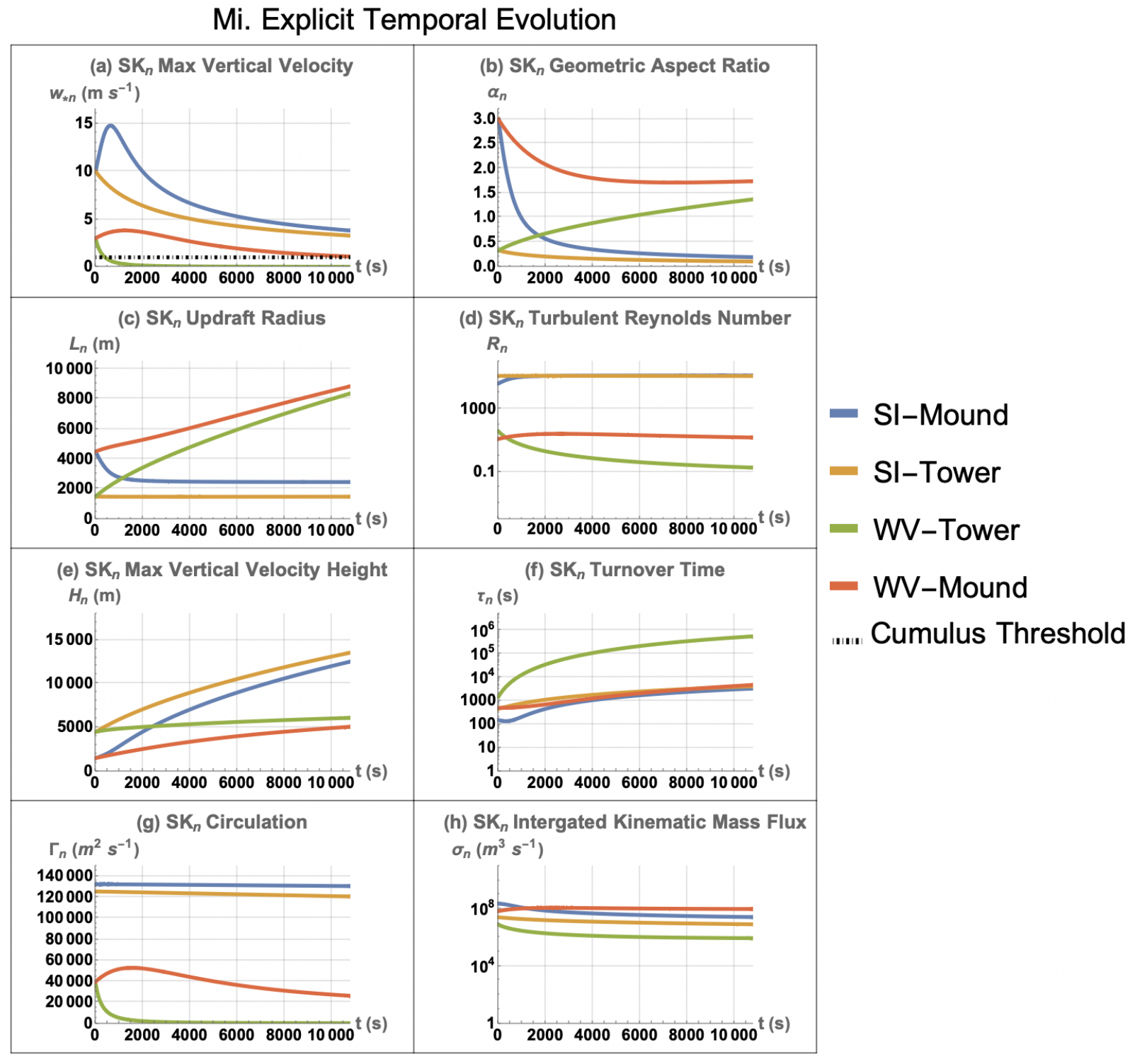}}
  \caption{SK explicit temporal evolution of implicit and explicit parameters and quantities of interest associated with the Mi. cloud-type test cases.}\label{Md Implicit Evolution}
\end{figure}

\begin{figure}[h]
 \centerline{\includegraphics[width=27pc]{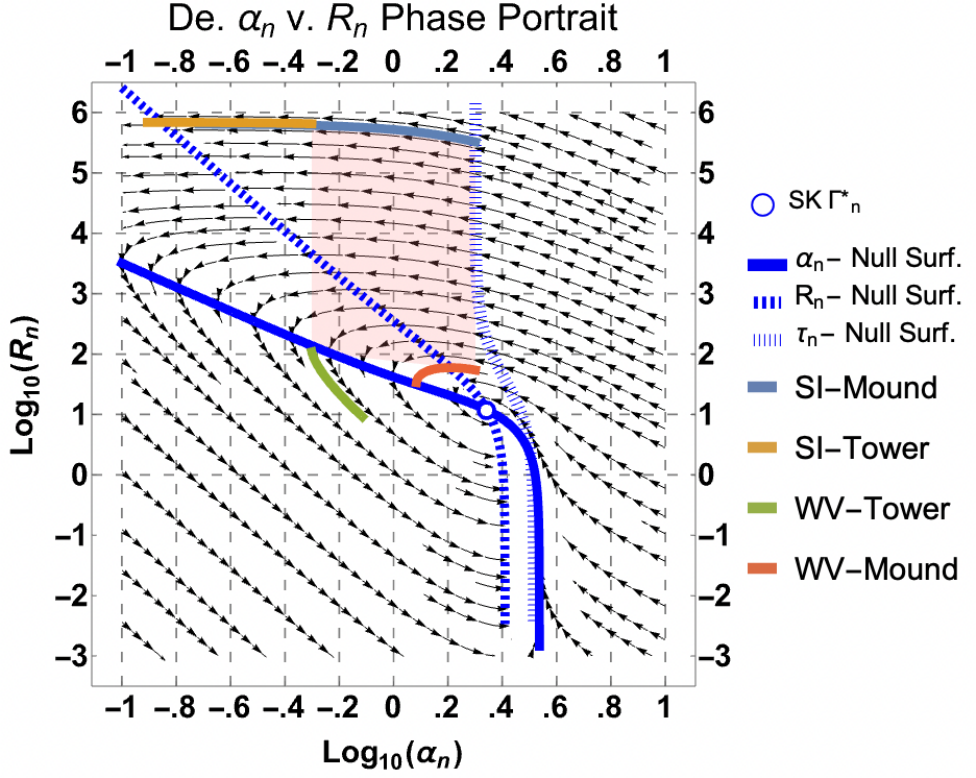}}
  \caption{Three-hour trajectory evolutions of the De. cloud type test cases projected onto the $\alpha_n$-$R_n$ subspace of the phase space. Here the pink-shaded quadrilateral corresponds to the $\alpha_n$-$R_n$ region of the phase space occupied by De. convection.}\label{Dp SK Phase Portrait}
\end{figure}

\begin{figure}[h]
 \centerline{\includegraphics[width=33pc]{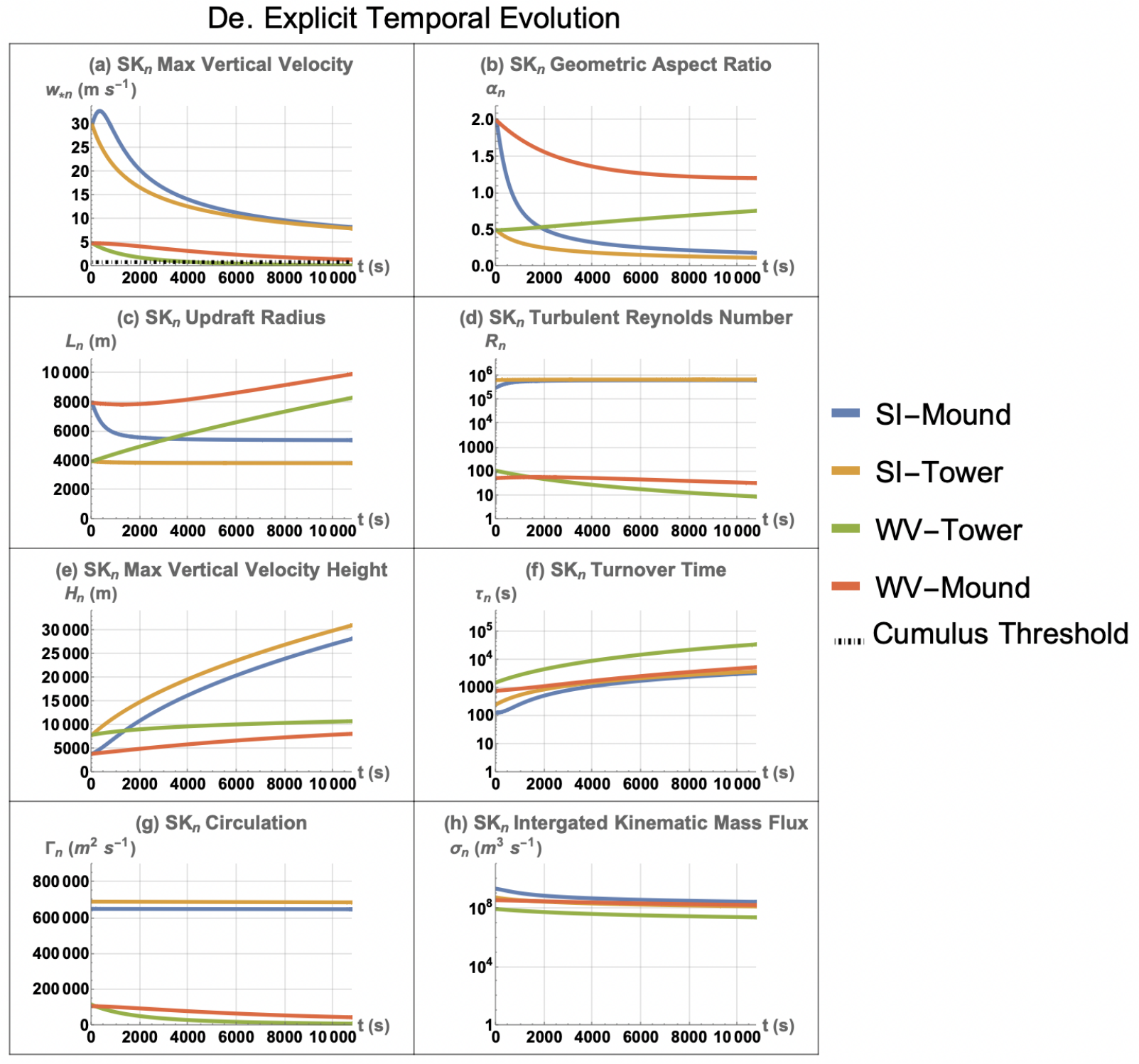}}
  \caption{SK explicit temporal evolution of implicit and explicit parameters and quantities of interest associated with the De. cloud-type test cases.}\label{Dp Implicit Evolution}
\end{figure}

Numerically integrating (\ref{SK Adv Dif Tend}) with the appropriate initial values, we investigate the temporal development of each test KRoNUT over a three-hour time interval. Although longer than the average lifetime described in Table \ref{Cloud Type}, this allows for a better evaluation of the long-term effects of cross-interaction once we transition to the $DK$ discussion.  The development of these SK test cases is then analyzed implicitly,  in the context of a phase portrait with test SK trajectories, (Fig. \ref{Sh SK Phase Portrait}, \ref{Md SK Phase Portrait}, and \ref{Dp SK Phase Portrait}), and explicitly by examining directly the temporal change in each parameter as well as the KRoNUT circulation and mass flux at the height of maximum vertical velocity, (Fig. \ref{Sh Implicit Evolution}, \ref{Md Implicit Evolution}, and \ref{Dp Implicit Evolution}) for the Sh., Mi., and De. cloud types respectively. 
\FloatBarrier
\subsubsection{Shallow Clouds}\label{subsection: Shallow Clouds}

\begin{figure}[h]
 \centerline{\includegraphics[width=27pc]{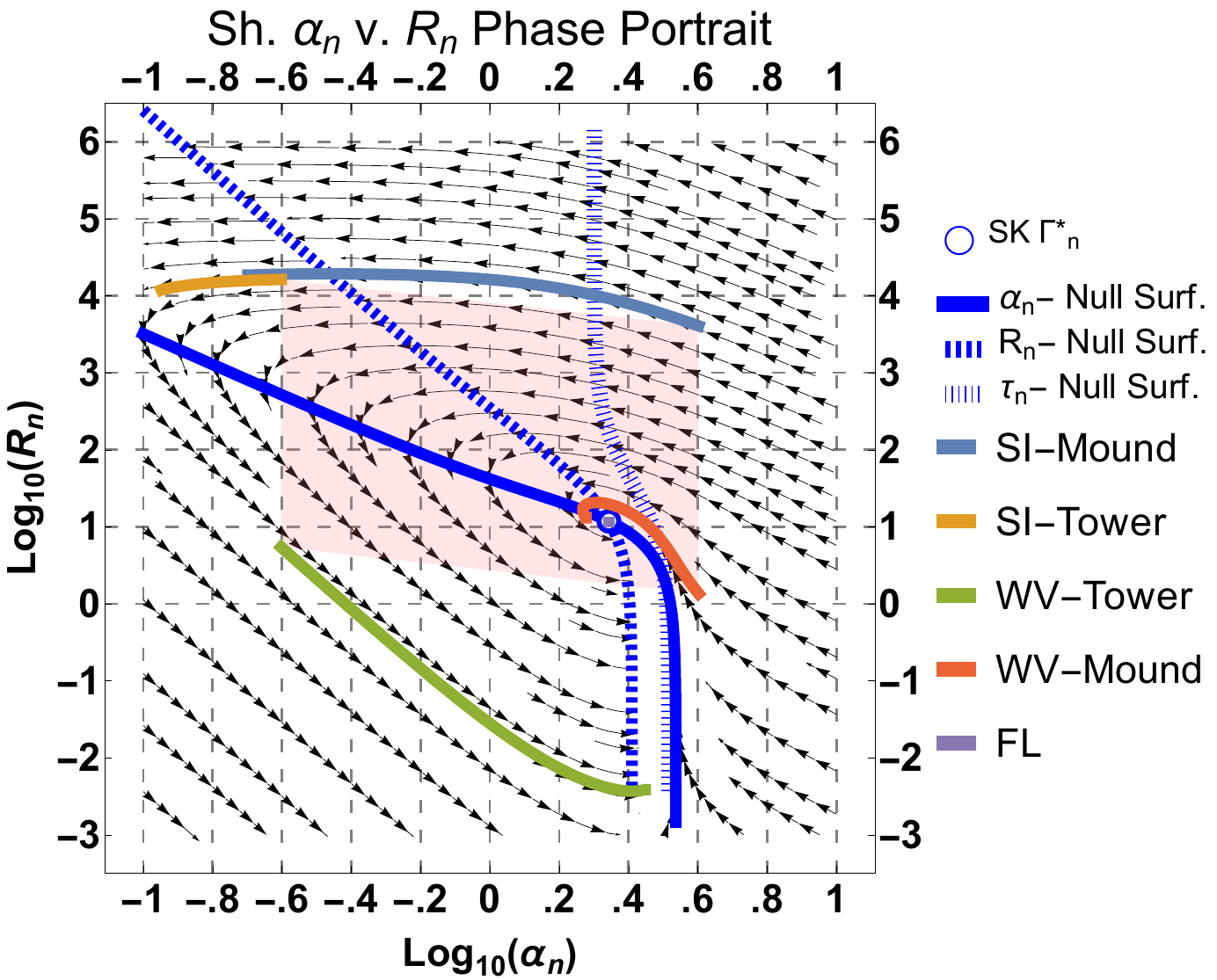}}
  \caption{Three-hour trajectory evolutions of the Sh. cloud type test cases projected onto the $\alpha_n$-$R_n$ subspace of the phase space.  Here the pink-shaded quadrilateral corresponds to the $\alpha_n$-$R_n$ region of the phase space occupied by Sh. convection.}\label{Sh SK Phase Portrait}
\end{figure}

\begin{figure}[h]
 \centerline{\includegraphics[width=33pc]{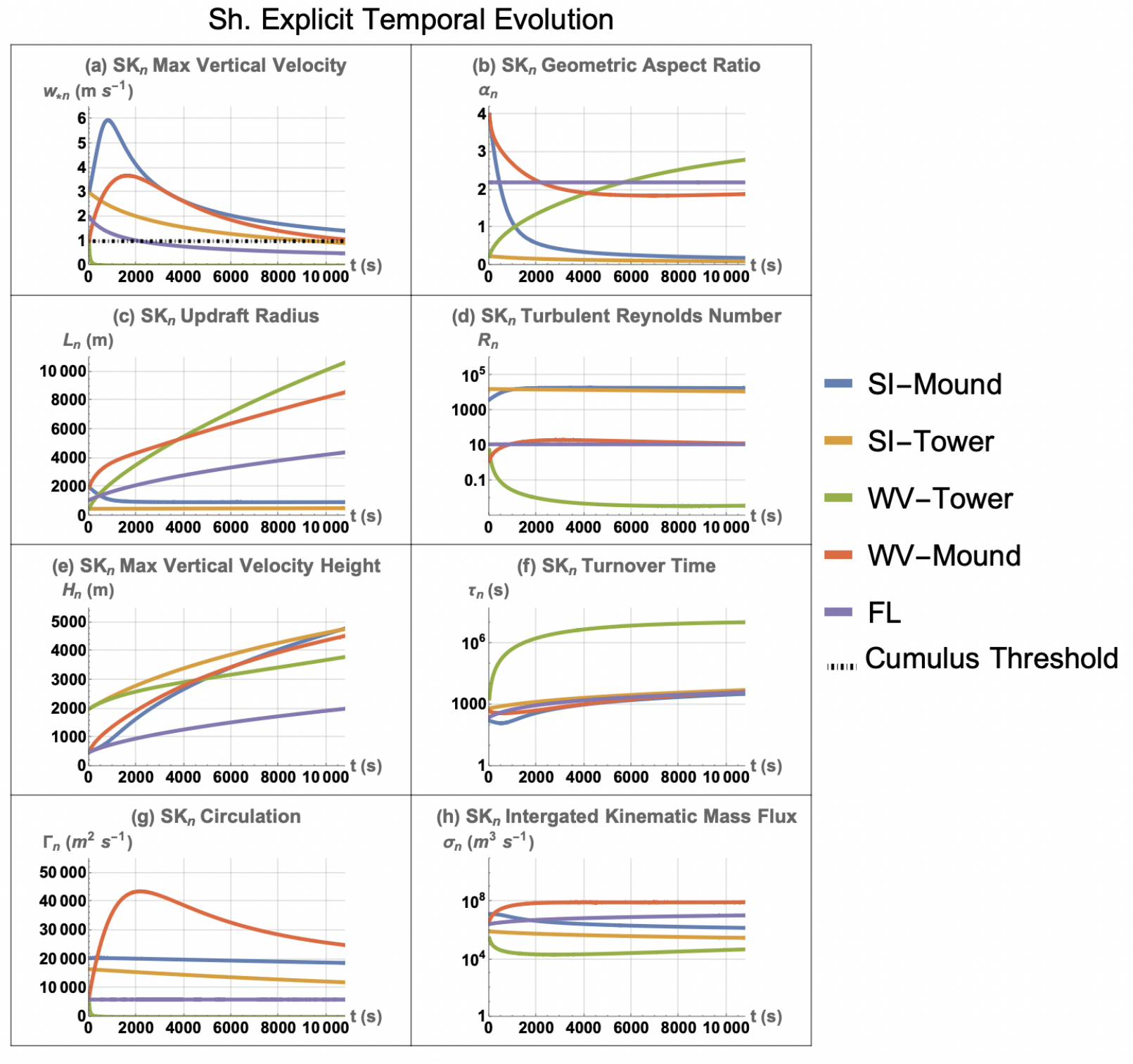}}
  \caption{SK explicit temporal evolution of implicit and explicit parameters and quantities of interest associated with the Sh. cloud-type test cases.}\label{Sh Implicit Evolution}
\end{figure}

We begin by analyzing the Sh. SI-Mound. As shown in Fig. \ref{Sh SK Phase Portrait}, this test SK does not remain in the Sh. regime. Rather, it initially invigorates its vertical velocity, thereby entering the mid-level regime, while also contracting its updraft radius in the process (Fig. \ref{Sh Implicit Evolution}). After this period of invigoration, which is associated with an increase in the Reynolds number, this SI-Mound goes into a state of enervation during which the contraction of the updraft radius and the change in Reynolds number levels off. This indicates that Sh. SI-Mounds prefer the state of an SI-Tower at which point the updraft radius and Reynolds number are in a quasi-steady state. As such, the increase in the height of maximum vertical velocity, which has a positive definite tendency, must be balanced by the decrease in maximum vertical velocity. As such, the Sh. SI-Tower exists in a state of slow decay, growing vertically while gradually weakening. 
\begin{table}[h]
\caption{List of test KRoNUTs that span the shallow cloud categories defined by (Table \ref{Cloud Type}). Test clouds are initialized so that they span each corner of the Sh. cloud type and include an additional test KRoNUT initialized at the fixed line.}\label{Shallow test cloud table}
\begin{center}
\begin{tabular}{|c|c|c|c|c|c|c|c|l|} 
  \hline
  Sh Type:& $\nu(m^2 s^{-1})$& $w_*(m s^{-1}$)& $L(m)$& $H(m$)& $\alpha $ & R  & $\tau (s)$\\\hline
  SI-Mound & 1.00& 3 & 2000& 500& 4.00& $4.08\times10^3$& 167\\\hline
  SI-Tower & 1.00& 3& 500 & 2000 & 0.250& $1.63\times10^4$& 667\\\hline
  WV-Tower & 1000& 1& 500& 2000& 0.250& 5.44& $2.00\times10^3$\\\hline
  WV-Mound & 1000 & 1 & 2000 & 500 & 4.00& 1.36& 500\\\hline
  FL & 229.48& 2& 1102.38 & 500 & 2.20& 11.8& 250\\\hline
\end{tabular}
\end{center}
\end{table} 
Contrasting this dwindling behavior to that of the Sh. WV-Tower, it becomes clear that the WV-Towers undergo a much more rapid decay. Unlike the quasi-steady state of the Sh. SI-Tower, the WV-Tower grows significantly in both the horizontal and the vertical causing, the maximum vertical velocity to plummet. As such the WV-Tower, unlike the SI-Mound and SI-Tower,  has a particularly volatile circulation and maximum vertical velcoity that rapidly decreases below the threshold for what we consider cumulus convection.  

Like the Sh. WV-Tower, the Sh. WV-Mound experiences a rapid change to its initial circulation; however, its evolution is markedly different. Although both experience an increase in updraft radius, rather than driving a rapid decrease in the maximum vertical velocity, in the case of the Sh. WV-Mound, this horizontal growth leads to steady invigoration. This strange behavior stems from the SK vector potential form, (\ref{general vector potential}), which has a vertical vorticity shear at the $z=0$ boundary that induces a spin-up of vorticity and thus circulation for certain geometries. For many SK cases, like that of the WV-Tower, it can be shown that this shearing acts as a vorticity sink term, enhancing decay; however, for wide geometries, $\alpha>2.205$, it serves as a vorticity source term. As such, Sh. WV-Mound experiences a phase in which it grows vertically and horizontally while increasing its maximum vertical velocity. This behavior increases the circulation, $\Gamma_n$, until  $\alpha_n$ decreases below 2.205 at which point the circulation slowly decays as the Sh. WV-Mound asymptotes toward the stable fixed circulation (Fig. \ref{Sh Implicit Evolution}). Finally, it's worth noting the behavior of Sh. FL.  Although Sh. FL remains at its fixed circulation, Fig. (\ref{Sh SK Phase Portrait}), its physical parameters, $w_{*n}$, $L_n$, and $H_n$, are perpetually evolving (Fig. \ref{Sh Implicit Evolution}).  It still grows upward and outward while weakening, but does so in such a way that its geometric aspect ratio, turbulent Reynolds number, and circulation remain constant (Fig. \ref{Sh Implicit Evolution} b,d, and g) respectively.
\FloatBarrier
\subsubsection{Mid-level Clouds}
Since the underlying structure of the phase space is fixed for the SK configuration, the differences between mid-level (congestus) circulations and shallow circulations are mostly magnitudinal, with some additional variation since we have initialized our test cases at different locations in the phase space based on observed cloud types.  One notable difference, however, is that given the range of parameters and the initialization of the test cases, Sh. WV-Mound rapidly becomes the strongest circulation of the Sh. test cases, whereas Mi. WV-Mound remains the third weakest circulation.  Additionally, it is interesting to note that despite variations in geometry and intensity, absolute invigoration between Sh. SI-Mound and Mi. SI-Mound as well as Sh. WV-Mound and Mi. WV-Mound are of similar magnitude. This would suggest that compared to their initial maximum vertical velocity, Sh. clouds can generate a similar magnitude of invigoration as their larger cloud counterparts. Finally, we note that the fixed line lies just outside the region we consider as mid-level clouds; thus, while clouds may asymptotically approach the fixed line, no Mi. clouds are initialized on it. As a final point, we acknowledge that the Mi. SI-Mound and SI-Tower reach unrealistic values of $H_n$ for the Earth's atmosphere during their lifecycle. This discrepancy is due to the absence of a tropopause in the current KRoNUT model.  Future work aims to explore how the evolution of these cloud-scale motions is altered by the presence tropopause.  

\begin{table}[h]
\caption{List of test KoNUTs that span the mid-level cloud categories defined by (Table \ref{Cloud Type}). Test clouds are initialized so that they span each corner of the Mi. cloud type.}\label{Mid test cloud table}
\begin{center}
\begin{tabular}{|c|c|c|c|c|c|c|c|l|} 
  \hline
  Md Type:& $\nu(m^2/s)$  & $w_*$(m/s) & $L$(m) & $H$(m) & $\alpha $ & R  & $\tau (s)$ \\\hline
  SI-Mound & 1 & 10 & 4500& 1500& 3.00& 150&$1.33\times10^5$\\\hline
  SI-Tower & 1& 10& 1500 & 4500 & 0.333& $1.22\times10^5$& 450\\\hline
  WV-Tower & 1000& 3& 1500& 4500&  0.333& 36.7& $1.50\times10^3$\\\hline
  MV-Mound & 1000 & 3 & 4500 & 1500 & 3.00& 12.2& 500 \\\hline
\end{tabular}
\end{center}
\end{table}

\FloatBarrier
\subsubsection{Deep Clouds}
Examining the deep cloud test cases, we observe that the cloud type echoes much of the behavior discussed in the previous subsection. Namely, while the magnitudes of parameter evolutions may change between cloud types, the underlying structure of the change is similar. Additionally, the De. test cases further support the notion that Sh. clouds can generate a similar magnitude of invigoration as their larger cloud counterparts. Much like the Mi. cases, De. suffer from a similar discrepancy in realistic $H_n$ values associated with the lack of a tropopause considered in the current form of our model. This analysis demonstrates that evolutions among shallow, mid-level, and deep cloud types are surprisingly alike; however, as shown in the following section, cloud types and their associated intensity and geometry become important as cross-interactions have the potential to alter the underlying structure of the phase space. 

\begin{table}[h]
\caption{List of test KRoNUTs that span the deep cloud categories defined by (Table \ref{Cloud Type}). Test clouds are initialized so that they span each corner of the De. cloud type.} \label{Deep cloud type table}
\begin{center}
\begin{tabular}{|c|c|c|c|c|c|c|c|l|} 
  \hline
  Dp Type:& $\nu(m^2/s)$  & $w_*$(m/s) & $L$(m) & $H$(m) & $\alpha $ & R  & $\tau (s)$ \\\hline
  SI-Mound & 1 & 30 & 8000 & 4000 & 2.00 & $3.26\times10^5$ & 133 \\\hline
  SI-Tower & 1 & 30 & 4000 & 8000 & 0.500& $6.52\times10^5$& 267 \\\hline
  WV-Tower & 1000& 5 & 4000& 8000& 2 & $\approx 5.44\times10^4$ & 800 \\\hline
  WV-Mound & 1000 & 5& 8000& 4000& 2 & $\approx 326.19$ & $\approx 133.33$\\\hline
\end{tabular}
\end{center}
\end{table}

\FloatBarrier 

\subsection{Double KRoNUT Dynamics}\label{subsection: Double DoNUT}
When considering two interacting KRoNUTs, the breakdown of rotational symmetry in the 
$DK$ configuration complicates how we treat the new cross-advection and deformation terms, see Appendix B for the full mathematical details. The essence, however, is as follows. First, we note that (\ref{conserved quantity budget}) holds in the vicinity of a KRoNUT$_n$'s convective core center, thus, we seek knowledge of KRoNUT$_f$'s velocity field about KRoNUT$_n$'s convective core center line. We arrive at this information by recasting the far-field KRoNUT's velocity in terms of the near field coordinate system and then applying a Taylor series expansion of the far field KRoNUT with regard to the near field radial variable, that is, about $r=0$. With this approximate velocity field, we then have all of the necessary pieces to treat the full form of (\ref{conserved quantity budget}). Similar to our $SK$ formulation, we finally equate the appropriate moments as defined by, (\ref{general equivalence of moments}), to arrive at the $DK$ dynamical system,
\begin{subequations}\label{DK parameter tendencies I}
    \begin{align}
        \begin{split}
            \dot{w}_{*n}=&-\frac{\text{w}_{*f} \text{w}_{*n}\text{H}_f  \left(\text{L}_f^2-(x_f-x_n)^2-(y_f-y_n)^2\right) e^{1-\frac{(x_f-x_n)^2+(y_f-y_n)^2}{\text{L}_f^2}}}{4 \text{L}_f^2 (\text{H}_f+\text{H}_n)^3 \left(8 \text{H}_n^3+3 \text{H}_n \text{L}_n^2\right)}\left(32 \text{H}_f \text{H}_n^3-4 \text{H}_n \text{L}_n^2 (3 \text{H}_f+4 \text{H}_n)-\text{L}_n^4\right)\\&+(\ref{SK Adv Dif Tend}a)
        \end{split}
        \\
        \begin{split}
            \dot{L}_n=&-\frac{\text{w}_{*f}\text{H}_f \text{L}_n  \left(\text{L}_f^2-(x_f-x_n)^2-(y_f-y_n)^2\right) e^{1-\frac{(x_f-x_n)^2+(y_f-y_n)^2}{\text{L}_f^2}}}{2 \text{L}_f^2 (\text{H}_f+\text{H}_n)^3 \left(8 \text{H}_n^2+3 \text{L}_n^2\right)}\left(-4 \text{H}_f \text{H}_n^2+3 \text{H}_f \text{L}_n^2+12 \text{H}_n^3+7 \text{H}_n \text{L}_n^2\right)\\&+(\ref{SK Adv Dif Tend}b)
        \end{split}
        \\
        \begin{split}
            \dot{H}_n=&\frac{\text{w}_{*f} \text{H}_f  \left(\text{L}_f^2-(x_f-x_n)^2-(y_f-y_n)^2\right)e^{1-\frac{(x_f-x_n)^2+(y_f-y_n)^2}{\text{L}_f^2}} }{4 \text{L}_f^2 (\text{H}_f+\text{H}_n)^3 \left(8 \text{H}_n^2+3 \text{L}_n^2\right)}\left(32 \text{H}_f \text{H}_n^3-4 \text{H}_n^2 \text{L}_n^2+\text{L}_n^4\right)\\&+(\ref{SK Adv Dif Tend}c)
        \end{split}
        \\
        \begin{split}
           \dot{x}_n=&\frac{\text{H}_f \text{w}_{*f} (x_f-x_n) e^{1-\frac{(x_f-x_n)^2+(y_f-y_n)^2}{\text{L}_f^2}}}{8 \text{L}_f^4 (\text{H}_f+\text{H}_n)^3 \left(4 \text{H}_n^2+\text{L}_n^2\right)} \Bigg\{\Bigg.3\text{H}_n \left(2 \text{L}_f^4 \text{L}_n^2-6 \text{L}_f^2 \text{L}_n^4+3 \text{L}_n^4 \left((x_f-x_n)^2+(y_f-y_n)^2\right)\right)+\\& \text{H}_f\left(8 \text{H}_n^2 \text{L}_f^4+2 \text{L}_f^4 \text{L}_n^2-6 \text{L}_f^2 \text{L}_n^4+3 \text{L}_n^4 \left((x_f-x_n)^2+(y_f-y_n)^2\right)\right)-8 \text{H}_n^3 \text{L}_f^4 \Bigg.\Bigg\}
        \end{split}
    \end{align}
\end{subequations}
Like the SK case, the tendency equations for the DK configuration are not especially useful in their basic form. To simplify this complexity, we once again make use of the implicit parameters, (\ref{dimensionless quantities}), while also introducing one new dimensional parameter and five new dimensionless parameters, 
\begin{subequations}\label{cross-dimensionless quantities}
    \begin{align}
        d&=\left(\left(x_f-x_n\right)^2+\left(y_f-y_n\right)^2\right)^{1/2}\label{DoNUT separation}\\
        \phi_{nf}&=\arctan\left(\frac{y_f-y_n}{x_f-x_n}\right)\label{separtion angle}\\
        \Lambda_f&=\frac{d}{L_f}\label{separation to Far field updraft radius ratio}\\
        \Pi_{nf}&=\frac{\text{H}_n}{\text{H}_f}\label{DK height of maximum vertical velocity ratio}\\
        \Upsilon_{nf}&=\frac{L_n}{L_f}\label{DK updraft radius ratio}\\
        \Xi_{nf}&=\frac{\text{w}_{*n}}{\text{w}_{*f}}\label{DK velocity ratio}
    \end{align}
\end{subequations}
where $d$ represents the separation between the convective cores of the two KRoNUTs, $\phi_{nf}$ corresponds to the angle of the separation as measured from the x-direction of the near field KRoNUT, $\Lambda_f$  identifies the ratio of the separation to the horizontal extent of the far field KRoNUT's updraft, and $\Pi_{nf}$,  $\Upsilon_{nf}$, and $\Xi_{nf}$ represent the ratio of the near field height of maximum vertical velocity, convective core updraft radius, and speed of maximum vertical velocity to their far field counterparts. Here, $d$ carries the units of meters while the other quantities are dimensionless. 

Using (\ref{cross-dimensionless quantities}) to rewrite (\ref{DK parameter tendencies I}) we find,
\begin{subequations}\label{DK parameter tendencies}
    \begin{align}
    \begin{split}
    \dot{w}_{*n}=&\frac{w_{*f}}{\left(3\alpha_n^2+8\right)}\Bigg\{\Bigg.\frac{\beth_{nf}}{\tau_n}\Bigg[\Bigg.-\frac{\Pi_{nf}}{4}\alpha_n^4-\left(4\Pi_{nf}+3\right)\alpha_n^2+8\Bigg.\Bigg]\Bigg.\Bigg\}+(\ref{parameter alpha Reynolds number tendency}a)
    \end{split}\label{DK wn non dimen param ten}
    \\
    \begin{split}
    \dot{L}_n=&\frac{w_{*f}}{\left(3\alpha_n^2+8\right)}\Bigg\{\beth_{nf}\Bigg.\Bigg[\Bigg.\left(\frac{7}{2}\Pi_{nf}+\frac{3}{2}\right)\alpha_n^3+\left(-2\Pi_{nf}+6\right)\alpha_n\Bigg.\Bigg]\Bigg.\Bigg\}+(\ref{parameter alpha Reynolds number tendency}b)
    \end{split}\label{DK Ln non dimen param ten}
    \\
    \begin{split}
    \dot{H}_n=&\frac{w_{*f}}{\left(3\alpha_n^2+8\right)}\Bigg\{\Bigg.\beth_{nf}\Bigg[\Bigg.-\frac{\alpha_n^4}{4}+\Pi_{nf}\alpha_n^2-8\Bigg.\Bigg]\Bigg.\Bigg\}+(\ref{parameter alpha Reynolds number tendency}c)
    \end{split}\label{DK Hn non dimen param ten}
    \\
    \begin{split}
    \dot{x}_n=&\frac{w_{*f}}{\left(\alpha_n^2+4\right)}\Bigg\{\Bigg.\frac{ \beth_{nf}\Lambda_f }{\left(\Lambda_f^2-1\right)}\Bigg[\Bigg.\frac{1}{\Upsilon_{nf}}\Bigg(\Bigg.\left(\frac{3}{4}\Pi_{nf}+\frac{1}{4}\right)\alpha_n^3+\left(-\Pi_{nf}+1\right)\alpha_n\Bigg.\Bigg)+\\& \Upsilon_{nf}\Bigg(\Bigg.\left(\Lambda_f^2-2\right)\left(\frac{9}{8}\Pi_{nf}+\frac{3}{8}\right)\alpha_n^3\Bigg.\Bigg)\Bigg.\Bigg]\Bigg.\Bigg\}
    \end{split}\label{DK xn non dimen param ten}
    \end{align}
\end{subequations}
where $\beth_{nf}=\frac{\left(\Lambda_f^{2}-1\right)e^{1-\Lambda_f^2}\Pi_{nf}}{\left(\Pi_{nf}+1\right)^3}$ is the cross-interaction coupling coefficient.  Here, we make use of the fact that Assumption \ref{assumption 3} implies $\dot{\phi}_{nf}=0$. Thus, without a loss of generality, we remove $\dot{y}_n$ from consideration by setting the initial separation angle to $\phi_{nf}=0$. In turn, the $\cos(\phi_{nf})$, which scales out front in $\dot{x}_n$, is equal to 1 while the $\sin(\phi_{nf})$ that scales of $\dot{y}_n$ is equal to 0 indicating that the whole term may be ignored. For further details, the curious reader is directed to Appendix C.

Importantly, (\ref{DK Ln non dimen param ten} a, b, c ) consist of terms resulting from cross-interaction as well as the self-interaction and diffusion given by (\ref{parameter alpha Reynolds number tendency}) while (\ref{DK xn non dimen param ten}) only results from cross-interactions.  While we offer an analysis of the terms in these equations, understanding how the system of equations evolves is considerably more convoluted. Examining (\ref{DK parameter tendencies}), we find that all cross-interaction terms  scale with $w_f$ and  that the cross-interaction terms are independent of $R_n$ or $R_f$. In most cases, the near field cloud experiences enervation due to cross interaction except where $\alpha_n$ or $\Pi_{nf}$ are very small. As for geometric growth, cross-interactions tend to enhance the near field updraft radius, although, in the $\lim_{\alpha_n\to0}$,  these effects tend toward zero. In the case of $H_n$, cross-interaction tends to suppress the near field height of maximum vertical velocity except for when $\Pi_{nf}$ is very large. Finally, turning our attention to $\dot{x}_n$, we observe $\dot{x}_n$ varies as $(\alpha_n^2+4)^{-1}$ as opposed to $(3\alpha_n^2+8)^{-1}$. Additionally, unlike the other terms, $\dot{x}_n$ depends on $\Upsilon_{nf}$ where the first term in square brackets scales with $\Upsilon_{nf}^{-1}$ and second scales with $\Upsilon_{nf}$. Thus, when $\Upsilon_{nf}\ll1$ the first term tends to dominate the behavior and vice versa for when $\Upsilon_{nf}\gg1$. In most cases, the first of these terms contributes to repulsion except when $\alpha_n$ is small and $\Pi_{nf}$ is large, in which case the term may become attractive. The second is entirely determined by $\Lambda_{nf}$ with $\Lambda_{nf}>2^{1/2}\approx1.41$ causing repulsion and $\Lambda_{nf}<2^{1/2}$ causing attraction.    

\begin{figure}[h]
 \centerline{\includegraphics[width=33pc]{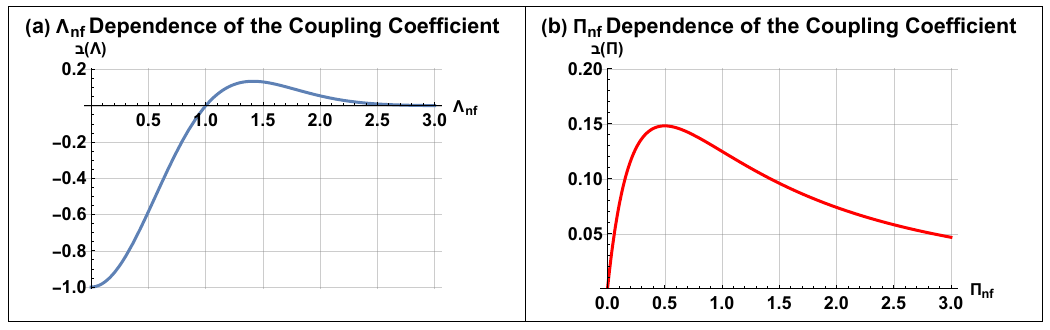}}
  \caption{The dependence of the coupling coefficient, $\beth_{nf}$ on $\Lambda_f$, panel (a) and $\Pi_{nf}$, panel (b) .}\label{beth_dependence}
\end{figure}

Common to the cross-interaction terms in $\dot{w}_{*n}$, $\dot{L}_{n}$, $\dot{H}_{n}$, and $\dot{x}_n$,   $\beth_{nf}$ exhibits non-monotonic behavior in both $\Lambda_f$ and $\Pi_{nf}$. Beginning with the $\Lambda_f$ dependence $\beth_{nf}$ is negative for $\Lambda_f<1$ and positive for $\Lambda_f>1$. This positive behavior in  $\Lambda_f$ peaks when $\Lambda_f=2^{1/2}\approx1.41$ before decaying exponentially to zero, (Fig. \ref{beth_dependence} a) . Given this rapid decay, cross-interactions quickly weaken with distance supporting the notion that cloud interactions are highly localized processes. Noting the well-separated assumption, Assumption \ref{assumption 2}, for simplicity, we limit further investigation to cases where $\Lambda_f\geq2$ initially, however, smaller values of $\Lambda_f$ are likely justifiable.  Turning our attention to the $\Pi_{nf}$ dependence, we find $\beth_{nf}$ is positive definite in $\Pi_{nf}$ with a peak in behavior occurring at $\Pi_{nf}=.5$ followed by a gradual decay to zero,  (Fig. \ref{beth_dependence} b). This behavior implies that, in terms of  $\beth_{nf}$, interactions are strongest when $H_f=2H_n$ although there is still a potential for strong interaction when $H_n\lessapprox H_f\lessapprox 2H_n$.  Together, these notions suggest that the potential for interesting interactions maximizes when circulations are nearby and when they are between clouds of the same type or a one-deeper type (i.e. shallow to shallow or mid-level to deep, for example). 

To further aid in our analysis, we also rewrite our implicit parameters given by (\ref{dimensionless quantities}) for the DK case,
\begin{subequations}\label{dimensionless tendencies double}
    \begin{align}
         \begin{split}
        \dot{\alpha}_n=\frac{1}{\left(3\alpha_n^2+8\right)}\Bigg\{\Bigg.\frac{\beth_{nf}}{\tau_f}\Bigg[\Bigg.\left(\frac{\Pi_{nf}}{4}\alpha_n^5+\left(\frac{5}{2}\Pi_{nf}+\frac{3}{2}\right)\alpha_n^3+\left(6\Pi_{nf}+6\right)\alpha_n\right)\Bigg]\Bigg.\Bigg\}\Bigg.+(\ref{dimensionless tendencies}a)\label{DK alpha dot}
        \end{split}
        \\
        \begin{split}
        \frac{\dot{R}_n}{R_n}=-\frac{1}{\left(3\alpha_n^2+8\right)}\Bigg\{\Bigg.\frac{\beth_{nf}}{\tau_f }\Bigg[\Bigg.\left(\frac{\alpha_n^4}{2}+\left(3\Pi_{nf}+3\right)\alpha_n^2\right)\Bigg]\Bigg.\Bigg\}\Bigg.+(\ref{dimensionless tendencies}b)\label{DK R dot}
        \end{split}
        \\
        \begin{split}
        \dot{\tau}_n=\frac{1}{\left(3\alpha_n^2+8\right)}\Bigg\{\Bigg.\frac{\beth_{nf}}{\Xi_{nf}}\Bigg[\Bigg.\left(\left(5\Pi_{nf}+3\right)\alpha_n^2-16\right)\Bigg]\Bigg.\Bigg\}\Bigg.+(\ref{dimensionless tendencies}c)\label{DK tau dot}
        \end{split}
    \end{align}
\end{subequations}
In this form, the tendencies due to cross-interactions are relatively straightforward functions that depend primarily on $\alpha_n$.  From ($\ref{DK alpha dot}$) we see that cross interactions always contribute to making $\dot{\alpha}_n$ more positive while (\ref{DK R dot}) shows that interactions always contribute to making $\dot{R}_n$ more negative. In both cases, the magnitude of these contributions increases with increasing $\Pi_{nf}$. By contrast, the sign of the cross-interaction term in $\dot{\tau}_n$ depends primarily on $\alpha_n$ with negative contributions occurring in tower-like geometries and positive contributions appearing in mound-like geometries. As $\Pi_{nf}$ increases, these contributions to $\tau_n$ are more likely to be positive. Unlike $\dot{\alpha}_n$ and $\dot{R}_n$, which scale with $\tau_f^{-1}$, $\dot{\tau}_n$ scales with $\Xi_{nf}^{-1}$. Thus, in the case of  $\dot{\alpha}_n$ and $\dot{R}_n$, cross interactions are strongest for far field KRoNUTs with short circulation times, that is, short KRoNUTs with strong maximum vertical velocities. By contrast,  $\dot{\tau}_n$'s  dependence on $\Xi_{nf}$ demonstrates that, for this quantity,  cross-interactions are strongest for far field KRoNUTs that have higher maximum vertical velocities than their near field counterparts.

Similar to the SK case, we now identify the location of the fixed line by setting $\dot{\alpha_n}=0$ and $\dot{R_n}=0$, and solving the associated system of equations for $\alpha_n^*$ and $R_n^*$ . Interestingly, the near field fixed line in the $DK$ configuration contains the same $\alpha_n^*$ as (\ref{SK fixed line}) however, $R_n^*$ now takes the form, 
\begin{equation}\label{DK fixed line}
\begin{split}
    R_n^*= \frac{16 \left(17 \sqrt{7}+32\right) e^{\Lambda_f ^2} \Xi_{nf}  (\Pi_{nf} +1)^3}{\left(\sqrt{7}+4\right) \left(\left(\sqrt{7}+13\right) e^{\Lambda_f ^2} \Xi_{nf}  (\Pi_{nf} +1)^3-16 \left(\Lambda_f ^2-1\right) \Pi_{nf}  \left(\left(2 \sqrt{7}+11\right) \Pi_{nf} +9\right)\right)}
\end{split}
\end{equation}
and, recalling (\ref{alpha Reynolds number circulation}), $\Gamma_n^*$ becomes   
\begin{equation}\label{DK preferred circulation}
    \Gamma_n^*=\nu \frac{16 \left(17 \sqrt{7}+32\right) e^{\Lambda_f ^2}  \Xi_{nf}  (\Pi_{nf} +1)^3}{3 \left(\sqrt{7}+13\right) e^{\Lambda_f ^2} \Xi_{nf}  (\Pi_{nf} +1)^3-48 \left(\Lambda_f ^2-1\right) \Pi_{nf}  \left(\left(2 \sqrt{7}+11\right) \Pi_{nf} +9\right)}
\end{equation}
Taking various limits of (\ref{DK fixed line}) and (\ref{DK preferred circulation}), see  (Table \ref{R_n^* limits}), we develop the following intuition.

\begin{table}[h]
\caption{Various limits of (\ref{DK fixed line}) and (\ref{DK preferred circulation})  for the $DK_n$ configuration. Importantly, $\Pi_{nf}=\frac{1}{\Pi_{fn}}$ and  $\Xi_{nf}=\frac{1}{\Xi_{fn}}$. Thus, case 3 for the near field implies case 4 for the far field and vice versa. Similarly, case 5 for the near field implies case 6 for the near field and vice versa.} \label{R_n^* limits}
\begin{center}

\begin{tabular}{|c|c|c|c|} \hline  
  
   Case:&Limit:& $R_n^*$: &$\Gamma_n^*$:\\ \hline 
   1&$\Lambda_f\rightarrow\infty$& $\frac{8}{27} \left(17 \sqrt{7}-5\right)$  &$\frac{16 \left(17 \sqrt{7}+32\right) \nu }{3 \left(\sqrt{7}+13\right)}$\\ \hline 
   2&$\Lambda_f\rightarrow 0$& $\frac{16 \left(17 \sqrt{7}+32\right) \Xi_{nf}  (\Pi_{nf} +1)^3}{\left(\sqrt{7}+4\right) \left(\left(\sqrt{7}+13\right) \Xi_{nf}  (\Pi_{nf} +1)^3+16 \Pi_{nf}  \left(\left(2 \sqrt{7}+11\right) \Pi +9\right)\right)}$  &$\frac{16 \left(17 \sqrt{7}+32\right)  \Xi_{nf}  (\Pi_{nf} +1)^3\nu}{3 \left(\sqrt{7}+13\right) \Xi_{nf}  (\Pi_{nf} +1)^3+48 \Pi_{nf}  \left(\left(2 \sqrt{7}+11\right) \Pi_{nf} +9\right)}$\\ \hline 
   3&$\Pi_{nf}\rightarrow\infty$& $\frac{8}{27} \left(17 \sqrt{7}-5\right)$  &$\frac{16 \left(17 \sqrt{7}+32\right) \nu }{3 \left(\sqrt{7}+13\right)}$\\ \hline 
   4&$\Pi_{nf}\rightarrow 0$ & $\frac{8}{27} \left(17 \sqrt{7}-5\right)$   &$\frac{16 \left(17 \sqrt{7}+32\right) \nu }{3 \left(\sqrt{7}+13\right)}$\\ \hline 
   5&$\Xi_{nf}\rightarrow\infty$&$\frac{8}{27} \left(17 \sqrt{7}-5\right)$  &$\frac{16 \left(17 \sqrt{7}+32\right) \nu }{3 \left(\sqrt{7}+13\right)}$\\ \hline 
   6&$\Xi_{nf}\rightarrow 0$ & 0  &0\\ \hline
\end{tabular}
\end{center}
\end{table}
In most cases, $R_n^*$ and $\Gamma_n^*$ approach the value associated with their $SK_n$ configuration, see (\ref{SK fixed line}), however, in cases 2 and 6, $R_n^*$ and $\Gamma_n^*$ approach new values. Case 2 corresponds to two clouds which have already merged. Although our well-separated assumption precludes a thorough numericalal investigation of this case, we may still speculate on the nature of this interaction. Given the geometry of a cloud-scale flow, we might expect that cloud overlap leads to a weakening of both circulations as the clouds approach merger. Furthermore, in this limit, one would expect that the filamentation of a cloud's vorticity decreases the circulation of an individual cloud. Given that Case 2 depends on $\Pi_{nf}$ and $\Xi_{nf}$ it is likely that the amount of circulation decrease depends on the ratios of the clouds' heights and intensities. If an additional limit corresponding to any of the Cases 3-6 is taken on Case 2,  $R_n^*$ and $\Gamma_n^*$ approach the values associated with their $SK_n$ configuration. Case 6 occurs when $w_n\rightarrow0$ or $w_f\rightarrow\infty$, however, unsurprisingly we only see cases where $w_n\rightarrow0$. As such, case 6 corresponds to the end of cloud$_n$'s cumulus form. Naturally, these changes in $R_n^*$ and $\Gamma_n^*$, as well as alterations to the null surfaces, alter the underlying structure of the phase space. To further develop our understanding of these behaviors,  we now turn to a numeric analysis of specific test cases.   

\subsection{Double KRoNUT Test Cases}
Fully evaluating the influence of cross-interaction is a large, multi-dimensional problem.   To determine potentially interesting configurations as well cases where the behavior remains nearly identical to their uncoupled SK counterparts, we let $DK_1$ be a particular test case defined in  (Tables \ref{Shallow test cloud table}, \ref{Mid test cloud table}, \ref{Deep cloud type table}). Allowing $DK_2$ to span a range of values associated with a particular cloud type (see Table \ref{Cloud Type}), we then examine which configurations of $DK_2$ alter the phase space of $DK_1$ and how these phase space changes affect the explicit temporal evolution of the KRoNUT parameters and other interesting quantities.  Notably, this analysis only identifies the subset of potentially interesting cases where the initial phase portrait is different due to the influence of a second circulation and misses cases where the effect may be significant later on in the cloud's evolution.  We now discuss two test particular cases which we believe of illustrative but far from exhaustive of potential interactions, one in which the interaction induces a strong change in the DK configuration compared to the SK counterpart, and one in which cross-interaction has a minimal effect.
\FloatBarrier
\subsubsection{DK Case 1}\label{subsubsection: DK Test 1}

We begin our $DK$ analysis by examining a case of strong interaction that results in the invigoration of one cloud at the cost of rapid enervation of its neighbor. This behavior occurs between the Sh. SV-Mound discussed in Section \ref{section: dynamical systems}(\ref{subsection: Shallow Clouds}) and a Sh. SV-Mound with initial conditions, $\nu=1000m^2 s^{-1}$,  $w_*=3 m s^{-1}$, $L=2000m$, $H=500m$, and $d=4000m$.  In this and all of the following section \ref{section: dynamical systems} subsections, we refer to parameters and quantities of interest in the context of their $DK$ form, with the $SK$ form being referenced where appropriate. To arrive at our qualitative understanding of the configuration, we quantitatively examine how both clouds simultaneously evolve from an implicit and explicit perspective. 

    \begin{figure}[h]
 \centerline{\includegraphics[width=27pc]{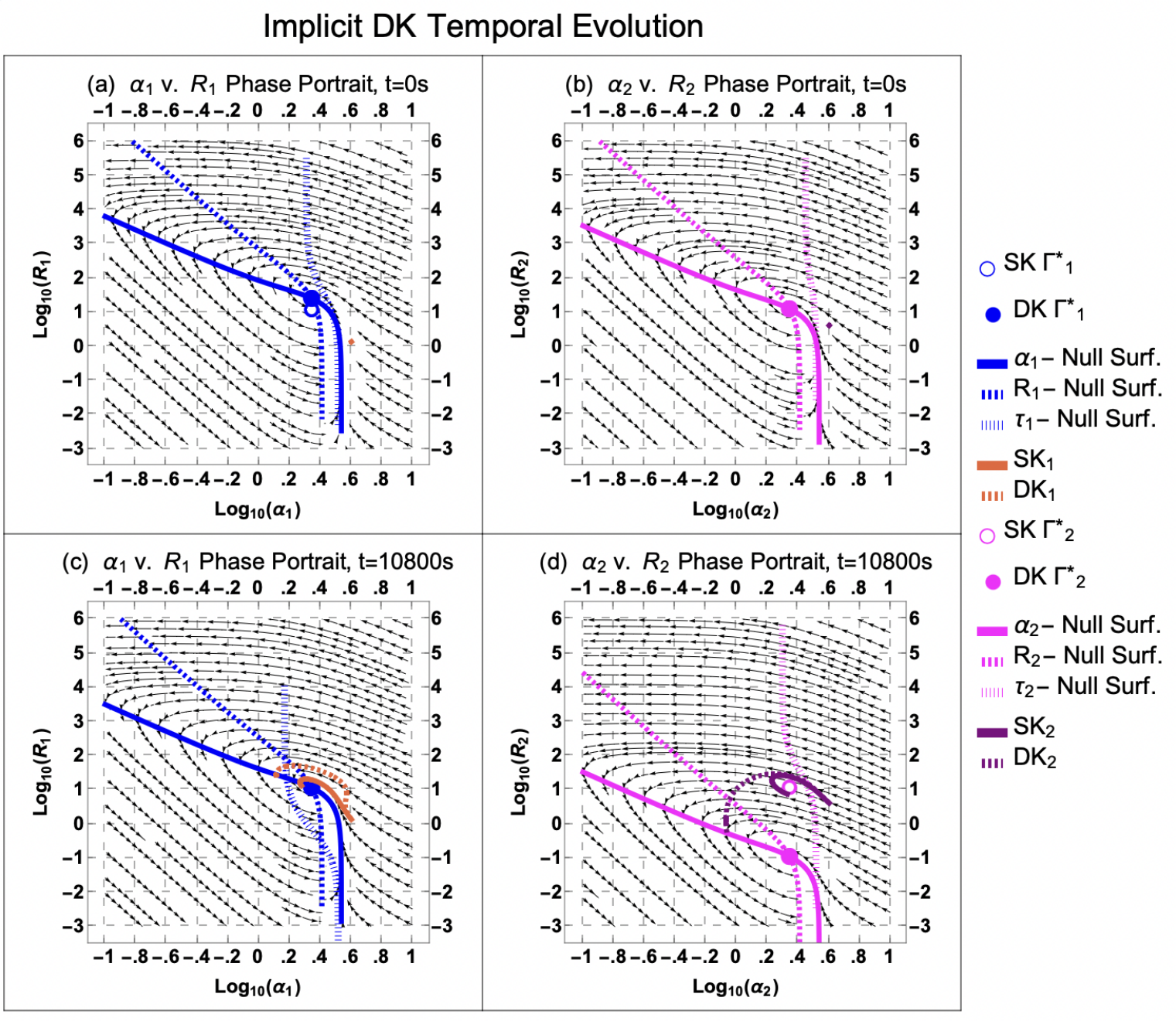}}
  \caption{Phase portraits of the $DK$ Case 1  interacting circulations. The left column shows portraits for $DK_1$. The right column shows portraits for $DK_2$.  The top row shows portraits at the initial time and the bottom row indicates the evolution at the final time.  Open circles mark the location of the attractive circulation in the SK case and filled circles for the DK case.}\label{Implicit evolution of DK case 1}
\end{figure}

   \begin{figure}[h]
 \centerline{\includegraphics[width=33pc]{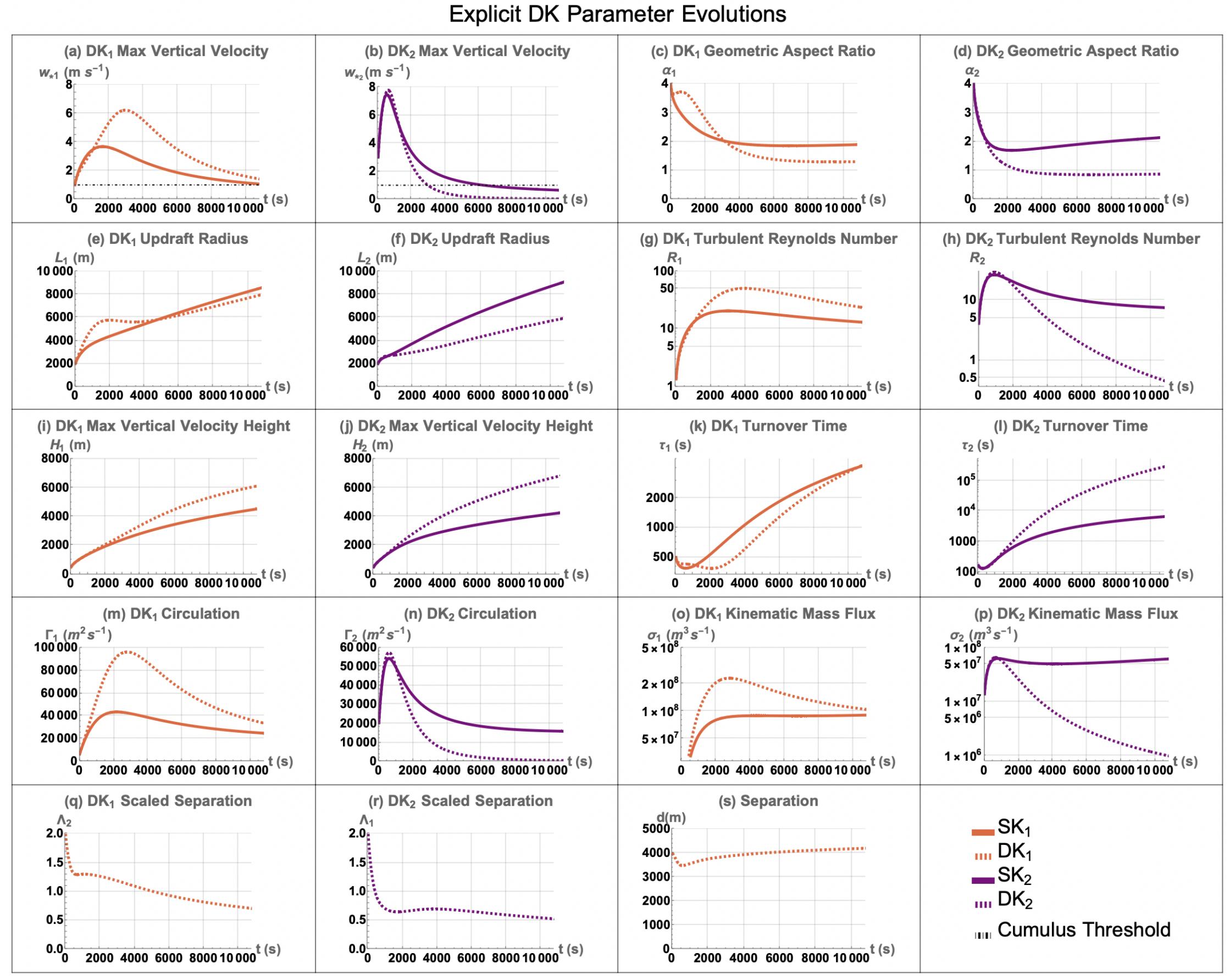}}
  \caption{DK Case 1 explicit temporal evolution of implicit and explicit KRoNUT parameters and their associated quantities of interest.}\label{Explicit evolution of DK case 1}
\end{figure}

  Starting from the implicit point of view, we examine the $DK_1$ and $DK_2$ phase portraits (Fig. \ref{Implicit evolution of DK case 1}), and observe that at the initial time, the values of $\Gamma_1^*$ and $\Gamma_2^*$ begin relatively close to their $SK$ configuration locations, with $\Gamma_1^*$ at a slightly higher Reynolds number (Fig. \ref{Implicit evolution of DK case 1}a and \ref{Implicit evolution of DK case 1}b). By the final time, however,  the $\Gamma_1^*$ gravitates toward its $SK$ placement while $\Gamma_2^*$ drifts toward $R_2=0$, thereby altering the trajectory of each cloud in the phase space (Fig. \ref{Implicit evolution of DK case 1}c and \ref{Implicit evolution of DK case 1}d). To better understand the implications of these alterations, we now turn our attention to the explicit temporal evolution of the $DK$ parameters and quantities of interest, (Fig. \ref{Explicit evolution of DK case 1}). 
   
   The explicit temporal evolution plots most intuitively illustrate the differences between $SK$ and $DK$ configurations. Given the exponential decay associated with $\Lambda_f$ in $\beth_{nf}$, we would expect cases of strong interaction to be those that reduce $\Lambda_f$ in time, as seen in (Figs. \ref{Explicit evolution of DK case 1}q and \ref{Explicit evolution of DK case 1}r). This behavior, however, does not necessarily imply that $d$ decreases for all time. Figure \ref{Explicit evolution of DK case 1}y and \ref{Explicit evolution of DK case 1}s, demonstrates a DK configuration that begins strongly attractive before becoming weakly repulsive. Despite the change of sign in $\dot{d}$, both $\dot{\Lambda}_{1}$ and $\dot{\Lambda}_2$, remain primarily negative due to the evolution in $L_1$ and $L_2$ (Fig. \ref{Explicit evolution of DK case 1}e and \ref{Explicit evolution of DK case 1}f). 
   
   We observe that the presence of $DK_2$ also induces a strong invigoration of the maximum vertical velocity, roughly $2m s^{-1}$ more than the $SK_1$ maximum, of $DK_1$ (Fig. \ref{Explicit evolution of DK case 1}a). Contrarily, although $DK_2$ reaches a slightly higher $w_*$ it then experiences rapid enervation causing a premature end to $DK_2$'s cumulus form (Fig. \ref{Explicit evolution of DK case 1}b). In turn, $R$,  $\Gamma^*$, and $\sigma$, all of which carry $w_*$ in their numerator, are enhanced for $DK_1$ (Fig. \ref{Explicit evolution of DK case 1}g, \ref{Explicit evolution of DK case 1}m, and \ref{Explicit evolution of DK case 1}o) and lessened for $DK_2$ (Fig. \ref{Explicit evolution of DK case 1}h, \ref{Explicit evolution of DK case 1}n, and \ref{Explicit evolution of DK case 1}p). It should be noted that despite the invigoration induced by the presence of $DK_2$, the time it takes $DK_1$ to reach the cumulus threshold is nearly identical to its $SK$ configuration, that is, the cloud still has approximately the same cumulus lifetime. In terms of geometry, we note that, although $\alpha_1$ goes through a period of fluctuation due to non-monotonic behavior in $L_1$, (Fig. \ref{Explicit evolution of DK case 1}e), both clouds settle into a more dome-like state, $\alpha\approx1$, than their $SK$ configurations (Fig. \ref{Explicit evolution of DK case 1}c and \ref{Explicit evolution of DK case 1}d). In both cases, the final dome-like structure is achieved by decreasing the final $L$ value of each cloud (Fig. \ref{Explicit evolution of DK case 1}e and \ref{Explicit evolution of DK case 1}f) while increasing their respective $H$ values (Fig. \ref{Explicit evolution of DK case 1}i and \ref{Explicit evolution of DK case 1}j). 
\FloatBarrier  
\subsubsection{DK Case 2}\label{subsubsection: DK Test 2}
Given the high intensity and relatively large geometries of Mi. and De. clouds, as well as our discussion of potentially strongly interacting clouds based on the DoNUT equations, one might expect a strong convective coupling between these cloud types. To this end, we examine the interaction between the Mi. SI-Tower discussed above, and a small Dp SI-Dome with initial conditions, $w_*=30 m s^{-1}$, $L=4000 m$, $H=4000 m$, and $d=8000 m$. We refer to these as $DK_1$ and $DK_2$ respectively. Although this particular configuration cannot be overly generalized, it does improve our understanding of Mi.-De. interaction as well as how SI-Towers evolve.      

    \begin{figure}[h]
 \centerline{\includegraphics[width=33pc]{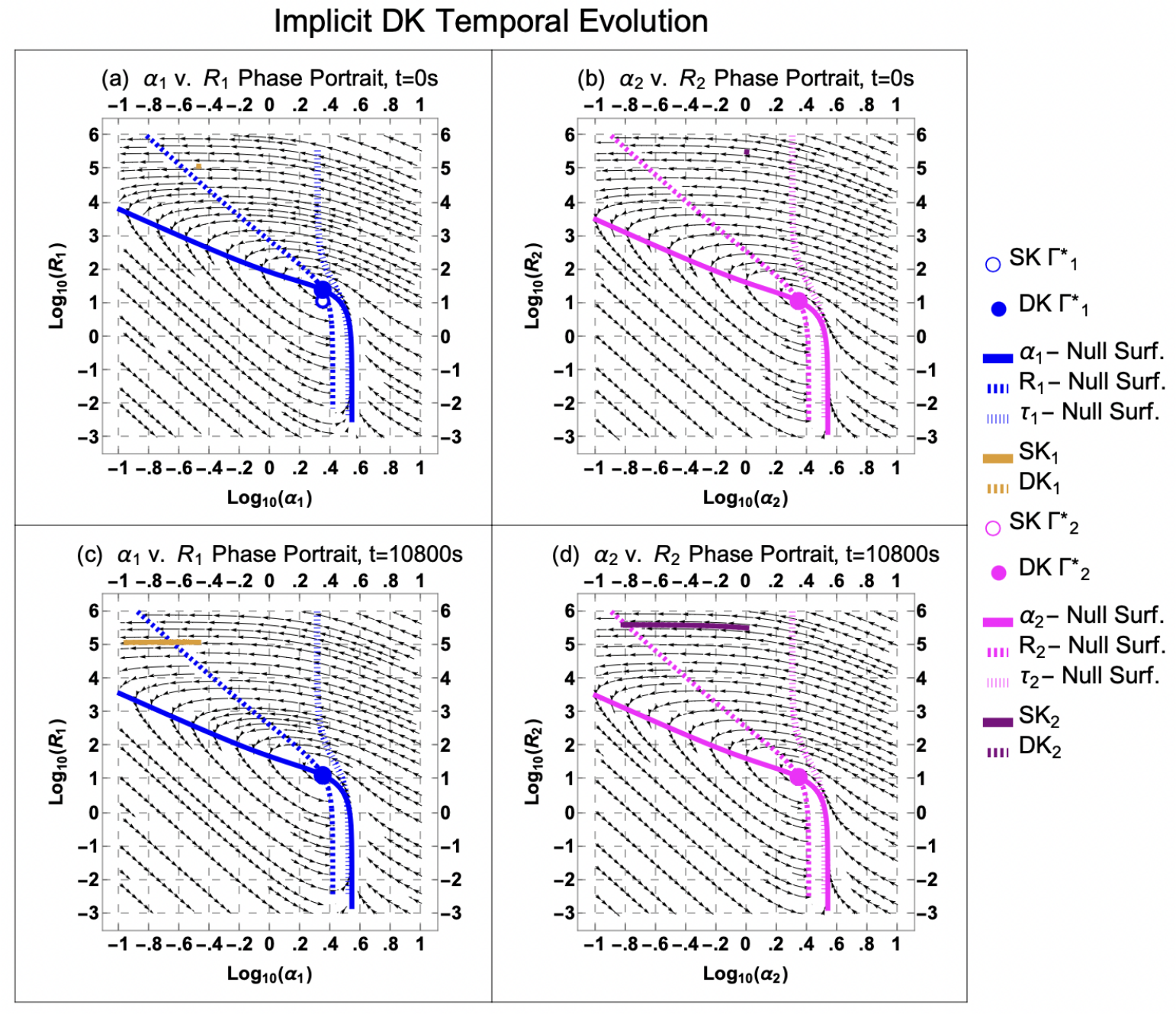}}
  \caption{Phase portraits of the two $DK$ Case 2 interacting circulations. The left column shows portraits for $DK_1$. The right column shows portraits for $DK_2$.  The top row shows portraits at the initial time and the bottom row, the final time.  Open circles mark the location of the fixed circulation in the SK case and filled circles correspond to the fixed circulation for the DK case.}\label{Implicit evolution of DK case 2}
\end{figure}

 \begin{figure}[h]
 \centerline{\includegraphics[width=39pc]{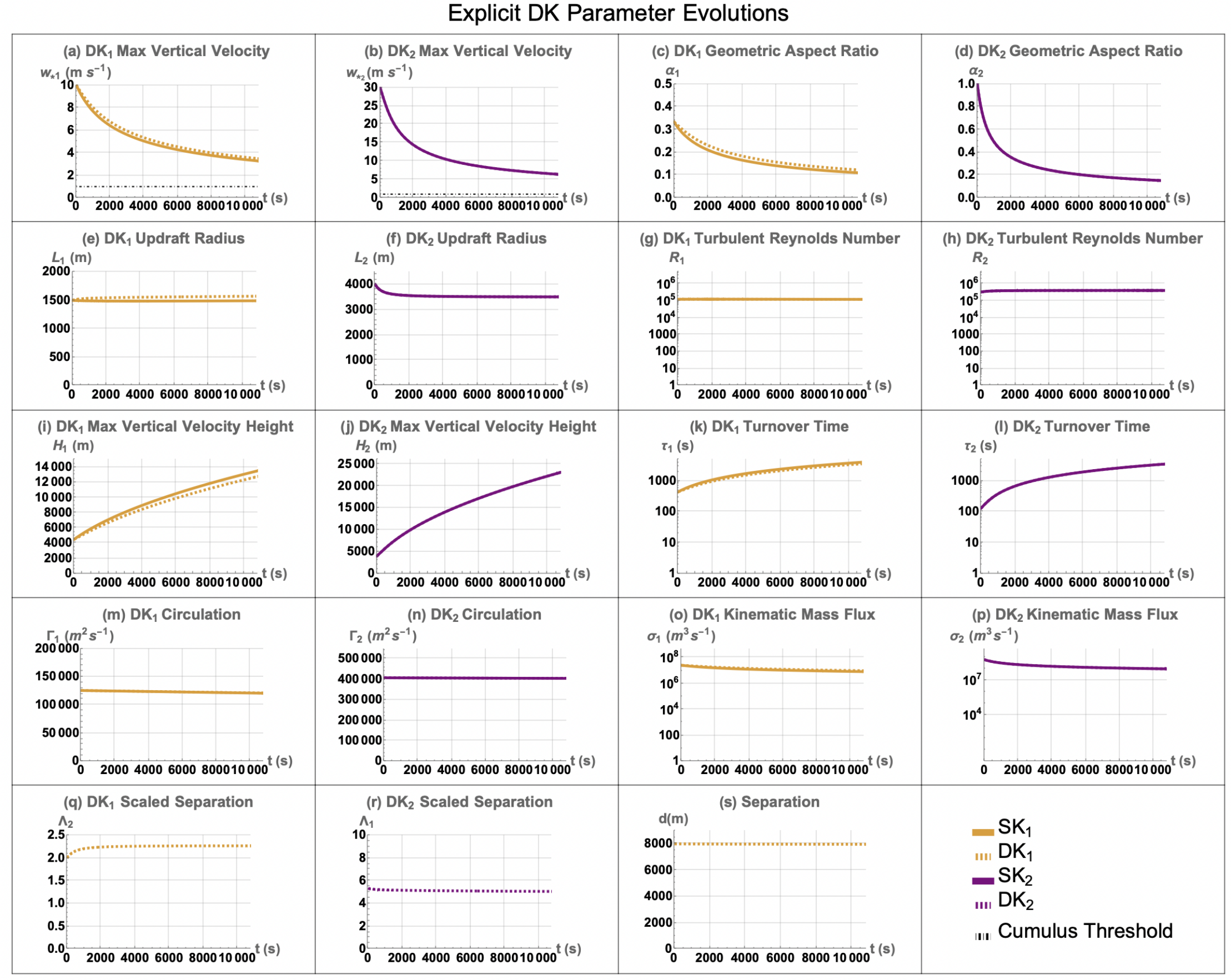}}
  \caption{DK Case 2 explicit temporal evolution of implicit and explicit KRoNUT parameters and their associated quantities of interest.}\label{Explicit evolution of DK case 2}
\end{figure}

Comparing (Fig. \ref{Implicit evolution of DK case 1}a and b) and (Fig. \ref{Implicit evolution of DK case 2}a and b), we observe that despite the differences in initial conditions, both configurations induce a similar effect on the initial phase planes. Namely, they lift $\Gamma_1^*$ to a higher Reynolds number. Despite this initial similarity, the final fixed circulation locations differ considerably between the two figures, (Fig. \ref{Implicit evolution of DK case 1}c and d) and (Fig. \ref{Implicit evolution of DK case 2}c and d). In $DK$ Case 1, we perceive a pronounced change in $\Gamma_2^*$ while $\Gamma_1^*$ approaches its $SK$ value. In $DK$ Case 2, however, $\Gamma_1^*$ approaches its $SK$ value while the location of $\Gamma_2^*$ remains largely unchanged by the cross interaction. Additionally, while the trajectories of $DK$ Case 1 drift away from their $SK$ forms, the trajectories of $DK$ Case 2 are indistinguishable from their $SK$ evolutions. As such, we expect that the explicit time evolutions of the parameters and the quantities of interest (Fig. \ref{Explicit evolution of DK case 2}) are largely similar to their $SK$ values. 

The explicit behavior illustrated in Fig. \ref{Explicit evolution of DK case 2} demonstrates the presence of weak coupling. This minimal interaction manifests in the $\Lambda_2$ and $\Lambda_1$ terms, (Fig. \ref{Explicit evolution of DK case 2} q and r). Unlike $DK$ Case 1 in which both clouds had the same horizontal size and thus $\Lambda$ values, the $DK$ Case 2 configuration sees different initial $\Lambda$ values with $\Lambda_2=2$ and $\Lambda_1=5.33$. As time progresses, $\Lambda_2$ increases before leveling off, thereby decreasing the potential for notable cross-interactions. By contrast, $\Lambda_1$ slowly decreases in time. However, its large initial $\Lambda$ value precludes it from reaching a low enough $\Lambda$ value to achieve strong coupling. These trends in $\Lambda$ stem from the associated behavior in $d$ and $L$. 

The separation between the clouds exists in a quasi-steady state, (Fig. \ref{Explicit evolution of DK case 2} s), thus all changes to $\Lambda$ are primarily attributed to changes in $L$ (Fig. \ref{Explicit evolution of DK case 2} e and f). In the case of $DK_1$, $L_2$ initially contracts before reaching a quasi-steady state, explaining the initial increase in $\Lambda_2$ followed by a period of little to no change. In contrast, although the interaction is weak, the presence of $DK_2$ causes a slight increase in $L_1$ in time. The growth of $DK_1$'s updraft radius causes $\Lambda_1$ to increase and $DK_2$ to feel $DK_1$ more strongly. However, the separation is still too great for any appreciable effect on the evolution. In addition to $\Lambda$, $\alpha$ also plays an important role in governing the behavior of the geometry of the circulations and thereby the potential for cross-interaction.

As demonstrated in (\ref{dimensionless tendencies double}), as $\alpha_n\rightarrow0$ so do the effects of cross-interaction in $\dot{\alpha}_n$ and $\dot{R}_n$. Thus, clouds that evolve toward a more tower-like state tend to experience $\alpha$ and $R$ evolutions that closely mirror their $SK$ evolutions. Given the minimal change in $L_1$ and $L_2$  and the large change in $H_1$ and $H_2$, it is unsurprising that the induced decrease in $\alpha_1$ and $\alpha_2$ leads to phase plane trajectories that nearly match their $SK$ evolutions.  

Examining $DK_2$ first, we note that there are no visual changes in the evolution however as previously mentioned DK$_1$ undergoes a sight increase in $L_1$ evolution due to the presence of $DK_2$. Additionally, $DK_2$ causes a small decrease in the $\dot{H}_1$ and a small increase in  $\dot{w}_{*i}$. Thus, in $DK_1$, although the effects of coupling are small, they act to slow the enervation due to self-interaction. Geometrically, they also reduce the contraction of the updraft radius associated with $SK$ behavior while suppressing the vertical growth of the cloud.  

\FloatBarrier
\section{Atmospheric Context and Discussion }\label{section: discussion}
The phase planes discussed in Subsection 5a  link the kinematics of the KRoNUT to the advection and diffusion dynamics of an individual cloud type. We took this analysis further by investigating how particular cases of cross-interaction alter these phase spaces and thus individual cloud evolutions, Subsection 5b. Building off this foundation, we now contextualize a KRoNUT's phase space by qualitatively considering a potential source of turbulent kinematic viscosity and atmospheric features that may alter a cloud's geometry. 

There are undoubtedly many factors that influence turbulent diffusion in the atmosphere, which is why we elect to analyze a wide range of turbulent kinematic viscosity, $\nu=1-1000m^2 s^{-1}$.  In the case of clouds, one of these potential physical mechanisms is related to spatial saturation gradients. The notion of spatial saturation gradients influencing a cloud's evolution is far from novel (\citet{prabhakaran_role_2020,holloway_moisture_2009,vraciu_role_2023}). In our case, we envision buoyancy fluctuations, induced by changes in saturation, as a mechanism to generate sub-cloud-scale eddies that readily diffuse the cloud-scale momentum. Thus, mixing of moist environmental air produces fewer buoyancy fluctuations, and thus fewer sub-cloud-scale eddies, than its dry air counterpart. Using this framework, we associate a correlation between moist environmental air and low turbulent kinematic viscosity. In the context of the $\alpha$-$R$-$\tau$  phase space, we imagine regions of high turbulent Reynolds number may occur in moist environmental air, potentially associated with cumulus preconditioning.  In contrast, we associate low turbulent Reynolds numbers with high kinematic viscosity induced by sharp moisture gradients between the cloud edge and the surrounding dry air.  In the case of the characteristic $SK$ phase plane, to the extent that a high turbulent Reynolds number may represent a high relative humidity environment, a moist ambient atmosphere could enable SI-Mounds to dynamically invigorate themselves as they evolve toward an SI-Tower state after which they slowly decay. By contrast, to the extent that a low turbulent Reynolds number can be associated with a low relative humidity, a WV-Tower in a dry environment undergoes rapid enervation as the momentum of its narrow convective core is readily dissipated to the environment. Interestingly, the WV-Mound, with its larger convective core, survives the rapid decay associated with high kinematic viscosity while asymptotically approaching the stable fixed circulation where the advection and diffusion forces balance as the KRoNUT slowly decays.  By contextualizing the turbulent kinematic viscosity in this way, we speculate that the saturation of the ambient could air play a critical role in determining the stable fixed circulation of a KRoNUT (\ref{SK preferred circulation}), and thus the evolution of the cumulus cloud-like flow.

 A complete qualitative evaluation of potential influences on the phase space requires a discussion of $\alpha$. To ground this conversation, we recall that this paper focuses on active clouds. That is, cumulus clouds that have reached their LFC, thereby gaining the ability to sustain their buoyancy through condensation. Before reaching this state, though, the cumulus cloud experience forced ascent driven by processes like thermals associated with sensible and latent heat release from the earth's surface and low-level convergence lines.  Thus, the initial geometry of active shallow clouds is determined by the geometry of the forced shallow cloud that comes before it. In the case of fair weather cumulus clouds, this geometry is often strongly influenced by the presence of a capping inversion, which acts as a barrier between the rising air in the mixed layer and the free troposphere above, causing the associated cumulus circulations to take on a ``pancake'' geometry, \citep{stull_fair-weather_1985}. Expanding this logic to the other cloud types, we note the potential for other layers that may limit the vertical growth of the cloud. In the case of the shallow to mid-level transition, one such layer would be the trade inversion, while in the case of mid-level to deep, it has been observed that the $T=0 K$ can act to inhibit vertical growth, \citep{johnson_trimodal_1999}. Furthermore, the tropopause could act as an additional barrier, hindering vertical growth and forcing clouds to grow outward, \citep{houze_cloud_2014}. Thus, these three layers provide conceptual mechanisms by which tower-like shallow, mid-level, or deep clouds may return to a more mound or dome-like geometry regardless of the underlying phase space.  Additionally, in the case of our model, they can provide a mechanism that obstructs the vertical growth of a particular circulation. This would be especially important for  De. cloud types, which, in the current simplified version of our model, grow to unrealistic heights (see De. cases 1 and 2).
 
\FloatBarrier
\section{Scientific Takeaways}
Before we delve into the scientific takeaways of this work, we once again draw attention to its scope. In this study, we conduct a series of relaxation experiments where we assume the existence of a pre-existing cloud field that has formed under the influence of baroclinicity, turbulent diffusion of its vorticity, and the advection and deformation of its vorticity associated with self and cross-cloud interactions. We then ``turn off'' the baroclinic term and investigate how combinations of the remaining terms influence the evolution of individual cloud-scale flows. Although the baroclinic term plays a fundamental role in cumulus cloud genesis, and over the entirety of the cloud's lifetime, by removing it from consideration, we can gain novel insights into how the remaining terms, which are often underappreciated, influence convective evolution and organization in their own right. These insights are as follows:

\begin{itemize}
     \item In the case of an isolated cloud, there exists a stable fixed circulation toward which all circulations evolve. To reach this stable circulation, diffusive clouds evolve along a particular path in an $\alpha_n$-$R_n$-$\tau_n$ phase space. In the case of our model, this path is determined by the initialization of the KRoNUT parameters, which we associate with the pre-existing cloud-scale flow at the moment that the baroclinic term is turned off. As cumulus cloud circulations evolve toward this stable fixed circulation, they eventually fail to maintain their cumulus form (i.e. a maximum vertical velocity greater than $1 m s^{-1}$) because of the enervation needed to reach the low Reynolds number associated with the fixed circulation. 
    \item  In the case of an isolated cloud, the specific value of the stable fixed circulation is determined by turbulent kinematic viscosity. 
    \item  The fixed circulation does not correspond to a steady-state convective mass flux through the height of maximum intensity, suggesting that the steady-state assumption used in plume model closures likely fails to capture the behavior of individual clouds.
    \item Isolated clouds can contract or expand, invigorate or enervate, but, in the absence of stable inversion layers, the clouds always grow vertically.  Thus, geometrically, upward growth of real clouds may be assumed, but nothing else may be.
    \item The rate of change of a cloud's height is not the same as the maximum vertical velocity. This suggests that clouds that grow vertically may also weaken, indicating that the cloud top rise rate can be a misleading proxy for airspeed in a cloud's convective core and should be used observationally with caution. 
    \item The presence of a neighboring cloud can alter the fixed circulation location and the underlying stability of the fixed circulation. 
    \item Clouds can act to push or pull each other closer or farther apart depending on their initial separation, geometry, and intensity.  This suggests that the three-dimensional structure of cumulus cloud circulations plays an important, non-trivial role in whether clouds are attracted or repelled from one another. 
    \item The potential for cloud coupling is set by the separation divided by the extent of a neighboring cloud's updraft radius. This suggests that to better appreciate the dynamic organization of cumulus convection, in addition to physical distance, the neighboring clouds' geometry, particularly their updraft radii, ought to be considered.
    \item The presence of a neighboring cloud does not necessarily lead to suppression of vertical growth or enervation of an arbitrary cloud, despite the potential for that cloud to exist in an exclusively subsiding environment.
\end{itemize}
As a final point on these takeways, although we have neglected baroclinicity in this study, there is no reason to assume that it, or any other additional term in the vorticity form of the Navier-Stokes conservation of momentum equation, would alter the current terms in our DoNUT equations as long as the function form of the current KRoNUT model stays the same. Instead, much like how turbulent diffusion and cross interactions appear as separate terms added to the self-interaction terms, it is likely that accounting for these additional processes will result in new terms being added to our preexisting DoNUT equations.  

\FloatBarrier
\section{Summary}\label{section: summary}

This paper investigates the dynamics associated with a kinematic representation of a pre-existing cloud-scale flow in the absence of baroclinicity. We effect this by stating a series of direct and tractable assumptions that allow us to simplify the vorticity budget equation. We then define the  Kinematic Representation of Non-rotating Updraft Tori (KRoNUT) model, which uses a low number of physically relevant, temporally dependent cloud-scale parameters,  to generate a poloidal velocity field that emulates kinematic features associated with a pre-existing cumulus cloud-scale flow. Using this model, we create a reduced-order framework via a moment reduction technique that mathematically approximates the dynamics resulting from our relaxation experiments. This technique yields the DoNUT equations, a dynamical system describing the evolution of cloud-scale parameters that govern the intensity, geometry, and location of the flow. This is done for the case of an isolated unit of convection, (\ref{parameter alpha Reynolds number tendency}), and for a pair of coupled convective units, (\ref{DK parameter tendencies}). Analyzing these dynamical equations, we develop an appreciation for how the physical cloud-scale quantities evolve in time and thus the temporal evolution of the associated cumulus cloud-scale circulation for shallow, mid-level, and deep morphologies. We then take this analysis a step further by providing cloud-relevant contextualization of the phase spaces associated with the isolated and coupled DoNUT equations. The findings derived from this methodology and subsequent analysis culminate in our scientific takeaways. 

As part of our scientific goal, we set out on a procedural objective - generate a framework that takes a particular KRoNUT representation of a cloud-scale motion and develops the associated DoNUT equations to explicate the temporal evolution of that flow due to the forcings in the momentum budget equation. Although we use a particular form of a KRoNUT and a limited set of forcings, the methodology developed in this paper is transferable to other KRoNUT functional forms and the inclusion of other physical processes that force a cumulus cloud's velocity field. This generality makes our framework adaptable to the inclusion of additional physics, such as baroclinicity associated with the thermodynamics of latent heating and cooling, radiative processes, shearing from background flows, and the Coriolis force. Additionally, it allows for the refinement of the functional form based on observational and computational fitting, as well as the inclusion of more than two convective units.  As such, in pursuit of our scientific aims, we have developed a general tool and procedure to aid in how we think about future parameterizations of convection. 

Historically, the convective community has emphasized the entraining plume model. Despite its usefulness, new developments in the field have become stagnant as its limited flow geometry and the steady-state assumption preclude it from the flexibility needed to treat all the processes associated with the cloud scale. As such, modern parameterizations likely neglect the potential upscale effects induced by cloud-scale interactions. In response to this deficiency, we introduce a new conceptual model, the KRoNUT, which sees a cloud-scale flow as the composite flow field generated by local and non-local motions associated with a series of thermals that comprise a thermal chain over a cumulus cloud's turnover timescale.  In doing so, we treat cumulus clouds as cohesive, 3-dimensional, time-dependent, fluid dynamic entities. The utility of this model is that by implementing a moment reduction technique, we achieve closed-form solutions for the cloud-scale evolutions while maintaining a tractable and relatively simple set of dynamical equations, DoNUTs. We thus develop the first leg of a new type of cloud parameterization which, although restricted by certain assumptions and the limited number of forcings considered, serves as the foundation for further innovation.  

%

%

\clearpage
\acknowledgments
This work is supported by NSF under Award AGS-2224293. D.F. was partially funded by Jastro-Shields Graduate Research Award.  J. Power provided help with the schematic.

%
%
\datastatement
The code used to generate the DoNUT equations and their subsequent analysis is available under the GitHub user DarioFalcone under the repository, Falcone et al. 2025. 

\appendix[A: Notation]
\label{AppendixA}

This section provides an overview of notation to help the curious reader follow the derivations and overall content of the paper. We will use a tilde to denote global variables, e.g. $\Tilde{x}$, $\Tilde{y}$, $\Tilde{z}$. In contrast, non-tilde variables denote the local variables associated with a near field KRoNUT, these include $x$, $y$, $z$, $r$, and $\theta$. The variable $z$ holds the same meaning for both the global and local cases, that is $\Tilde{z}=z$. Given that we primarily work in the coordinates of the near field KRoNUT, we elect to use $z$ for simplicity. A vector quantity, $\vec{\psi}$, $\vec{u}$, or $\vec{\omega}$, followed by a superscript, denotes the component of that vector. Superscripts on non-vector quantities indicate powers. A partial, $\partial$, followed by a subscript indicates a partial derivative with respect to the subscript variable. We refer to the time rate of change of a quantity as its tendency equation, with additional comments given on whether it is the spatial or material description when needed.  A dot over the quantity is reserved for the total temporal derivative of a quantity; for example, $\frac{d\alpha(t)}{dt}=\dot{\alpha}$.

Subscripts $g$, $n$, $f$, $1$, and $2$ indicate a KRoNUT and its associated parameters. Here, a subscript $g$ denotes a quantity of a general, arbitrary KRoNUT, $n$ designates an arbitrary near field KRoNUT, while $f$ indicates an arbitrary far field KRoNUT. The reason for these distinctions will become clear in Section 3 as we derive the near-field circulation density budget. Subscripts $1$ and $2$ signify a particular KRoNUT, namely $DK_1$ and $DK_2$. The near field action of KRoNUT$_1$ on itself plus the far field action of KRoNUT$_2$ on KRoNUT$_1$ is written DK$_{12}$.  While the same interactions on KRoNUT$_2$ are denoted  DK$_{21}$. A subscript $T$ denotes a total quantity. $\kappa$ represents the weight of a particular moment, thus $M_\kappa$ represents the $\kappa$ moment of the circulation density (i.e. $\gamma$), while $N_\kappa$ corresponds to the $\kappa$ moment of the circulation density equation (i.e. $\partial_t\gamma$).  To simplify equations in section \ref{section: dynamical systems}, we reference previous equation numbers within the equations that are introduced there.  
\begin{center}
Table 1A: Scripting notation

\begin{table}[h!]
\centering
\begin{tabular}{|c|c|}
 \hline
  Quantity & Definition \\\hline\hline
  $s_g$&(subscript) General quantity 
  \\\hline
  $s_n$&(subscript) Near field 
  \\\hline
  $s_f$&(subscript) Far field 
  \\\hline
  $s_{1,2}$&(subscript) A particular KRoNUT in the double KRoNUT configuration
  \\\hline
  $s_T$&(subscript) Total quantity 
  \\\hline
  $\Vec{s}^{x,r,etc.}$&(superscript) Vector component 
  \\\hline
  $\partial_{x,r,etc.}\vec{s}$& (subscript) Partial WRT spacial coordinate or time 
  \\\hline
  $\dot{s}$& Total time derivative
  \\\hline
  $s^*$& Denotation of a quantity associated with the fixed line in the phase space 
  \\\hline
\end{tabular}
\caption{Outline of how subscripts and superscripts are used throughout the paper.}
\label{table: Notation}
\end{table}
\end{center}
\FloatBarrier
\begin{center}
Table 2A: Quantities Used in the Paper

\begin{table}[h!]
\centering
\begin{tabular}{|c|c|}
 \hline
  Quantity & Definition \\
 &\\\hline
 $\Tilde{x}$&Global $x$ variable 
 \\\hline
  $\Tilde{y}$&Global $y$ variable 
 \\\hline
  $\Tilde{z}$&Global $z$ variable 
 \\\hline
  $x$&Local $x$ variable to the convective core of the near field KRoNUT
  \\\hline
  $y$&Local $y$ variable to the convective core of the near field KRoNUT
 \\\hline
 $z$&Local $z$ variable to the convective core of the near field KRoNUT. Note $z=\Tilde{z}.$
 \\\hline
  $\vec{\psi}_g$&Vector potential
  \\\hline
  $\vec{u}_g$&Velocity field
  \\\hline
  $\vec{\omega}_g$&Vorticity field
  \\\hline
  $\gamma_g$& Circulation density
  \\\hline
  $\Gamma_g$& Circulation 
  \\\hline
  $\nu$&Diffusion coefficient (global quantity)
  \\\hline
  $\kappa$&Weight associated with a particular moment 
  \\\hline
  $M_k$&The moment of the conserved quantity of the flow, $\gamma_n$\\\hline
  $N_k$&The moment of the local time rate of change for the conserved quantity of the flow, $\partial_t\gamma_n$\\\hline
  $\alpha_g$&Aspect ratio of geometric parameters, i.e the ratio of $\text{L}_g$ to $\text{H}_g$\\\hline
  $R_g$&Turbulent Reynolds number 
  \\\hline
  $\tau_g$&Circulation turnover time 
  \\\hline 
  $\alpha_g^*$&Aspect ratio associated with the fixed line 
  \\\hline
  $R_g^*$&Turbulent Reynolds number associated with the fixed line
  \\\hline
  $\Gamma_g^*$&Fixed circulation associated which occurs along the fixed line 
  \\\hline
 $d$&Separation between KRoNUT convective core centers \\\hline 
 $\phi_{nf}$&The angle of separation as measured from the near field DoNUT\\\hline 
 $\Lambda_f$&The ratio of the $d$ to $\text
{L}_f$  \\\hline 
 $\Pi_{nf}$&The ratio of the $\text{H}_n$ to $\text
{H}_f$ \\\hline 
 $\Upsilon_{nf}$&The ratio of the $L_n$ to $L_f$\\\hline
 $\Xi_{nf}$&The ratio of the $w_{*n}$ to $w_{*f}$\\\hline
 \end{tabular}
\caption{Key quantities and their definition}
\label{Quantities}
\end{table}
\end{center}
\FloatBarrier

\appendix[B]
To arrive at the cross interaction tendency equations, (\ref{DK parameter tendencies}), we first perform the variable transform associated with (\ref{p transform of variables}) but now on $\vec{\psi}_f$
\begin{subequations}
    \begin{align}
        \psi_f^x&=-\frac{w_{*f} z (-y_f+y_n+y) e^{\left(1-\frac{z}{\text{H}_f}-\frac{(-x_f+x_n+x)^2+(-y_f+y_n+y)^2}{\text{L}_f^2}\right)}}{2 \text{H}_f}\\
        \psi_f^y&=\frac{w_{*f} z (-x_f+x_n+x) e^{\left(1-\frac{z}{\text{H}_f}-\frac{(-x_f+x_n+x)^2+(-y_f+y_n+y)^2}{\text{L}_f^2}\right)}}{2 \text{H}_f}\\
        \psi_f^z&=0
    \end{align}
\end{subequations}
Then, once again applying a coordinate transform such that, $\vec{\psi}_f$ is described by a cylindrical coordinate system with its azimuthal angle about the z-axis and origin at $(x_n, y_n, 0)$, we rewrite KRoNUT$_f$'s vector potential as,
\begin{subequations}\label{o vector potential in cylindrical}
    \begin{align}
        \psi_f^r&=\frac{w_{*f} z \Bigg[\sin (\theta) (x_n-x_f)+\cos (\theta) (y_f-y_n)\Bigg]  e^{1-\frac{z}{\text{H}_f}-\frac{(-x_f+x_n+r \cos (\theta))^2}{\text{L}_f^2}-\frac{(-y_f+y_n+r \sin (\theta))^2}{\text{L}_f^2}}}{2 \text{H}_f}\\
        \psi_f^\theta&=\frac{w_{*f} z \Bigg[\cos (\theta) (x_n-x_f)+\sin (\theta) (y_n-y_f)+r\Bigg] e^{1-\frac{z}{\text{H}_f}-\frac{(-x_f+x_n+r \cos (\theta))^2}{\text{L}_f^2}-\frac{(-y_f+y_n+r \sin (\theta))^2}{\text{L}_f^2}}}{2 \text{H}_f}\\
        \psi_f^z&=0.
    \end{align}
\end{subequations}
To simplify the integration associated with the necessary moments, we now perform a Taylor series expansion of $r$ in $\vec{\psi}_f$ about the point $r=0$ out to the quadratic term which yields, 
\begin{subequations}\label{approximate far field vector potential}
    \begin{align}
        \begin{split}
            \psi_f^r\approx&\frac{\text{w}_{*f} z e^{1-\frac{z}{\text{H}_f}-\frac{(x_f-x_n)^2+(y_f-y_n)^2}{\text{L}_f^2}}}{2 \text{H}_f \text{L}_f^4} \Bigg[\sin (\theta) (x_n-x_f)+\cos (\theta) (y_f-y_n)\Bigg]\\&\Bigg\{\Bigg.\text{L}_f^4+r \Bigg[\Bigg.2 \cos (\theta) (x_f-x_n) \left(\text{L}_f^2+2 r \sin (\theta) (y_f-y_n)\right)-\left(r \cos ^2(\theta) \left(\text{L}_f^2-2 (x_f-x_n)^2\right)\right)+\\&\sin (\theta) \left(2 \text{L}_f^2 (y_f-y_n)-r \sin (\theta) \left(\text{L}_f^2-2 (y_f-y_n)^2\right)\right)\Bigg.\Bigg]\Bigg.\Bigg\}
        \end{split}\\
        \begin{split}
            \psi_f^\theta\approx&\frac{\text{w}_{*f} z e^{1-\frac{z}{\text{H}_f}-\frac{(x_f-x_n)^2+(y_f-y_n)^2}{\text{L}_f^2}}}{2 \text{H}_f \text{L}_f^4} \Bigg\{\Bigg.\text{L}_f^4 r-(\cos (\theta) (x_f-x_n)+\sin (\theta) (y_f-y_n))\\& \Bigg[\Bigg.\text{L}_f^4+r^2 \left(-3 \text{L}_f^2+(x_f-x_n)^2+(y_f-y_n)^2\right)+ r \Bigg(\Bigg.2 \cos (\theta) (x_f-x_n) \left(\text{L}_f^2+2 r \sin (\theta) (y_f-y_n)\right)+\\&2 \text{L}_f^2 \sin (\theta) (y_f-y_n)+r \cos (2 \theta) (x_f-x_n+y_f-y_n) (x_f-x_n-y_f+y_n)\Bigg.\Bigg)\Bigg.\Bigg]\Bigg.\Bigg\}
        \end{split}
        \\
        \begin{split}
           \psi_f^z=&0.
        \end{split}
    \end{align}
\end{subequations}
Note that the vector potential now has a component in the $r$ direction and the $\theta$ direction, both of which carry a dependence on $\Tilde{\theta}_n$. Applying (\ref{velocity vector from potential}) to (\ref{approximate far field vector potential}), we arrive at the approximate far field velocity,
\begin{subequations}\label{approximate far field velocity}
    \begin{align}
        \begin{split}
            u_f^r&\approx\frac{\text{w}_{*f} (z-\text{H}_f) e^{1-\frac{z}{\text{H}_f}-\frac{(x_f-x_n)^2+(y_f-y_n)^2}{\text{L}_f^2}}}{2 \text{H}_f^2 \text{L}_f^4} \Bigg\{\Bigg.\text{L}_f^4 r-\Bigg(\cos (\theta) (x_f-x_n)+\sin (\theta) (y_f-y_n)\Bigg)\\& \Bigg[\Bigg.\text{L}_f^4+r^2 \left(-3 \text{L}_f^2+(x_f-x_n)^2+(y_f-y_n)^2\right)+r \Bigg(\Bigg.(2 \cos (\theta) (x_f-x_n) \left(\text{L}_f^2+2 r \sin (\theta) (y_f-y_n)\right)+\\&2 \text{L}_f^2 \sin (\theta) (y_f-y_n)+r \cos (2 \theta) (x_f-x_n+y_f-y_n) (x_f-x_n-y_f+y_n)\Bigg.\Bigg)\Bigg.\Bigg]\Bigg.\Bigg\}
        \end{split}
        \\
        \begin{split}
            u_f^\theta&\approx\frac{\text{w}_{*f} (\text{H}_f-z) e^{1-\frac{z}{\text{H}_f}-\frac{(x_f-x_n)^2+(y_f-y_n)^2}{\text{L}_f^2}}}{2 \text{H}_f^2 \text{L}_f^4} \Bigg[\sin (\theta) (x_n-x_f)+\cos (\theta) (y_f-y_n)\Bigg]\\& \Bigg\{\Bigg.\text{L}_f^4+r \Bigg[\Bigg.2 \cos (\theta) (x_f-x_n) \left(\text{L}_f^2+2 r \sin (\theta) (y_f-y_n)\right)-\left(r \cos ^2(\theta) \left(\text{L}_f^2-2 (x_f-x_n)^2\right)\right)+\\&\sin (\theta) \left(2 \text{L}_f^2 (y_f-y_n)-r \sin (\theta) \left(\text{L}_f^2-2 (y_f-y_n)^2\right)\right)\Bigg.\Bigg]\Bigg.\Bigg\}
        \end{split}
        \\
        \begin{split}
            u_f^z&\approx\frac{\text{w}_{*f} z e^{1-\frac{z}{\text{H}_f}-\frac{(x_f-x_n)^2+(y_f-y_n)^2}{\text{L}_f^2}}}{\text{H}_f \text{L}_f^4} \Bigg\{\Bigg.\text{L}_f^2 \left(\text{L}_f^2-(x_f-x_n)^2-(y_f-y_n)^2\right)-\\&2 r \left(-2 \text{L}_f^2+(x_f-x_n)^2+(y_f-y_n)^2\right) \left(\cos (\theta) (x_f-x_n)+\sin (\theta) (y_f-y_n)\right)\Bigg.\Bigg\}
        \end{split}     .
    \end{align}
\end{subequations}

We now have all of the necessary pieces to compute all of the DK terms in (\ref{general equivalence of moments}). Taking the appropriate DK moments, we note that there are several similarities between them and their SK counterparts. Thus, in an attempt to simplify DK system equations, we use previous numbered equations from the SK system.  It should be understood that with this notation, which we use throughout this subsection, we mean the right-hand side of the referenced equation. The moments thereby take the form,
\begin{subequations}\label{DK moment budgets}
    \begin{align}
        \begin{split}
            N_{r^0}=&(\ref{diffusive SK budget moment r0})\label{DK Mr^0t}
        \end{split}
        \\
        \begin{split}
            N_{\Tilde{x}^1 }=&\frac{\text{H}_f \text{w}_{*f} \text{w}_{*n} (x_f-x_n) e^{2-\frac{(x_f-x_n)^2+(y_f-y_n)^2}{\text{L}_f^2}}}{32 \text{H}_n \text{L}_f^4 (\text{H}_f+\text{H}_n)^3} \Bigg\{\Bigg.-8 \text{H}_n^3 \text{L}_f^4+\\&\text{H}_f \Bigg[\Bigg.8 \text{H}_n^2 \text{L}_f^4+2 \text{L}_f^4 \text{L}_n^2-6 \text{L}_f^2 \text{L}_n^4+3 \text{L}_n^4 \left((x_f-x_n)^2+(y_f-y_n)^2\right)\Bigg.\Bigg]+\\&3 \text{H}_n \Bigg[\Bigg.2 \text{L}_f^4 \text{L}_n^2-6 \text{L}_f^2 \text{L}_n^4+3 \text{L}_n^4 \left((x_f-x_n)^2+(y_f-y_n)^2\right)\Bigg.\Bigg]\Bigg.\Bigg\}-\frac{e \nu \text{w}_{*n} x_n \left(32 \text{H}_n^4+8 \text{H}_n^2 \text{L}_n^2-3 \text{L}_n^4\right)}{4 \text{H}_n^3 \text{L}_n^2}\label{DK Mx^1t}
        \end{split}
        \\
        \begin{split}
             N_{\Tilde{y}^1 }=& \frac{\text{H}_f \text{w}_{*f} \text{w}_{*n} (y_f-y_n) e^{2-\frac{(x_f-x_n)^2+(y_f-y_n)^2}{\text{L}_f^2}}}{32 \text{H}_n \text{L}_f^4 (\text{H}_f+\text{H}_n)^3} \Bigg\{\Bigg.-8 \text{H}_n^3 \text{L}_f^4+\\&\text{H}_f \Bigg[\Bigg.8 \text{H}_n^2 \text{L}_f^4+2 \text{L}_f^4 \text{L}_n^2-6 \text{L}_f^2 \text{L}_n^4+3 \text{L}_n^4 \left((x_f-x_n)^2+(y_f-y_n)^2\right)\Bigg.\Bigg]+\\&3 \text{H}_n \Bigg[\Bigg.2 \text{L}_f^4 \text{L}_n^2-6 \text{L}_f^2 \text{L}_n^4+3 \text{L}_n^4 \left((x_f-x_n)^2+(y_f-y_n)^2\right)\Bigg.\Bigg]\Bigg.\Bigg\}-\frac{e \nu \text{w}_{*n} y_n \left(32 \text{H}_n^4+8 \text{H}_n^2 \text{L}_n^2-3 \text{L}_n^4\right)}{4 \text{H}_n^3 \text{L}_n^2}\label{DK My^1t}
        \end{split} 
        \\
        \begin{split}
             N_{z^1 }=&-\frac{\text{H}_f \text{H}_n \text{w}_{*f} \text{w}_{*n} \left(4 \text{H}_f \text{H}_n+\text{L}_n^2\right) e^{2-\frac{(x_f-x_n)^2+(y_f-y_n)^2}{\text{L}_f^2}} \left(-\text{L}_f^2+(x_f-x_n)^2+(y_f-y_n)^2\right)}{2 \text{L}_f^2 (\text{H}_f+\text{H}_n)^3}\\&+(\ref{diffusive SK budget moment z1})\label{DK Mz^1t}
        \end{split}
        \\
        \begin{split}
            N_{r^2}=&-\frac{\text{H}_f \text{L}_n^4 \text{w}_{*f} \text{w}_{*n} (\text{H}_f+3 \text{H}_n) e^{2-\frac{(x_f-x_n)^2+(y_f-y_n)^2}{\text{L}_f^2}} \left(\text{L}_f^2-(x_f-x_n)^2-(y_f-y_n)^2\right)}{4 \text{H}_n \text{L}_f^2 (\text{H}_f+\text{H}_n)^3}\\&+(\ref{diffusive SK budget moment r2})\label{DK Mr^2t}
        \end{split}
    \end{align}
\end{subequations}
Finally, we equate the moments of (\ref{DK moment budgets}) to the appropriate time derivatives of the moments unlined in  (\ref{circulation density moments}). Rearranging these equations we arrive at the $DK$ DoNUT equations given by (\ref{DK parameter tendencies}).

\appendix[C]
To further reduce (\ref{DK parameter tendencies}), we demonstrate the presence of another conserved quantity, namely $\phi_{nf}=0$. Noting the similarity in the form of $\dot{x}_n$ and $\dot{y}_n$ we elect to recast $x_n$ and $y_n$ so that the location of the convective core center is described in a polar coordinate with its origin located at the $\Gamma_T$ centroid. As such we get the following relations,
\begin{subequations}\label{polar centroid relations}
    \begin{align}
         x_n&=r_{nf}\cos{\phi_{nf}}\\
         y_n&=r_{nf}\sin{\phi_{nf}}
    \end{align}
\end{subequations}
where $r_{nf}$ is the radial distance from the $\vec{\Gamma}_T$ centroid. Taking a time derivative of (\ref{polar centroid relations}) we find, 
\begin{subequations}\label{general x_n and y_n dot equations}
    \begin{align}
        \dot{x}_n&=\dot{r}_{nf}\cos{\phi_{nf}}-\dot{\phi}_{nf}r_{nf}\sin{\phi_{nf}}\label{x_n dot centroid relation}\\
        \dot{y}_n&=\dot{r}_{nf}\sin{\phi_{nf}}+\dot{\phi}_{nf}r_{nf}\cos{\phi_{nf}}\label{y_n dot centroid relation}.
    \end{align}
\end{subequations}
Equating (\ref{DK xn non dimen param ten}) to (\ref{x_n dot centroid relation}) and the $y_n$ form of (\ref{x_n dot centroid relation}) to (\ref{y_n dot centroid relation}) while dividing the by $\cos{\phi}_{nf}$ and $\sin{\phi}_{nf}$, respectively, we arrive at the following relations, 
\begin{subequations}
    \begin{align}
        \aleph_{nf}=\dot{r}_{nf}-r_{nf}\tan{\phi_{nf}}\dot{\phi}_{nf}\label{aleph from x dot}\\
        \aleph_{nf}=\dot{r}_{nf}+r_{nf}\cot{\phi_{nf}}\dot{\phi}_{nf}\label{aleph from y dot}
    \end{align}
\end{subequations}
Equating (\ref{aleph from x dot}) and (\ref{aleph from y dot}) and solving for $\dot{\phi}_{nf}$ one finds that,
\begin{equation}\label{phi dot}
    \dot{\phi}_{nf}=0
\end{equation}
which is a direct result of Assumption 2 and constitutes the second conserved quantity of the system. Using (\ref{DK xn non dimen param ten}), (\ref{x_n dot centroid relation}), and (\ref{phi dot}) we now write the near field centroid distance tendency as,
\begin{equation}\label{centroid distance dot}
    \dot{r}_{nf}=\aleph_{nf}.
\end{equation} 
Making use of notation associated with a specific DK we note, 
\begin{equation}\label{d in terms of centroid distance}
    d=r_{ij}+r_{ji}.
\end{equation}
Taking a time derivative of (\ref{d in terms of centroid distance}) while making use of (\ref{centroid distance dot}) one finds, 
\begin{equation}\label{d dot}
    \dot{d}=\aleph_{ij}+\aleph_{ji}.
\end{equation}
Making the reduction associated with (\ref{phi dot}
) and (\ref{d dot}), we have shown that the four tendencies, $\dot{x}_1$, $\dot{y}_1$, $\dot{x}_2$, and $\dot{y}_2$, may be reduced to a single tendency equation, $\dot{d}$. Thus, the tenth-order system of the DK configuration may be reduced to a seventh-order system under Assumption 3.






%




\bibliographystyle{ametsocV6}

\bibliography{references}
\end{document}